\documentclass[journal]{IEEEtran}
\usepackage{cite}
\usepackage{amsmath,amssymb,amsfonts}
\usepackage{algorithmic}
\usepackage{graphicx}
\usepackage{optidef}
\usepackage{enumitem} 
\usepackage{float}
\usepackage{svg}  
\usepackage{float} 
\usepackage{placeins} 
\usepackage[ruled,vlined]{algorithm2e}
\usepackage{amsmath}

\usepackage{textcomp}
\usepackage{xcolor}
\usepackage{subcaption}
\usepackage{bm}
\def\BibTeX{{\rm B\kern-.05em{\sc i\kern-.025em b}\kern-.08em
    T\kern-.1667em\lower.7ex\hbox{E}\kern-.125emX}}

\ifCLASSINFOpdf
\else
\fi
\begin{document}
%
\title{Bridging the Modality Gap: Enhancing Channel Prediction with Semantically Aligned LLMs and Knowledge Distillation}
%
%
%
\author{Zhaoyang Li,~\IEEEmembership{Graduate Student Member, IEEE}, Qianqian Yang,~\IEEEmembership{Member, IEEE}, \\ Zehui Xiong,~\IEEEmembership{Senior Member, IEEE},   Zhiguo Shi and Tony Q. S. Quek,~\IEEEmembership{Fellow, IEEE}
\IEEEcompsocitemizethanks{
\IEEEcompsocthanksitem Zhaoyang Li, and Qianqian Yang, and Zhiguo Shi are with the College of Information Science and Electronic Engineering, Zhejiang University, Hangzhou,
China. (e-mails: \{zhaoyangli, qianqianyang20, shizg\}@zju.edu.cn).
\IEEEcompsocthanksitem Zehui Xiong and Tony Q. S. Quek are with the Singapore University of Technology and Design. (e-mails: \{zehui\_xiong, tonyquek\}@sutd.edu.sg). 
}
}

\markboth{Journal of \LaTeX\ Class Files,~Vol.~14, No.~8, August~2015}%
{Shell \MakeLowercase{\textit{et al.}}: Bare Demo of IEEEtran.cls for IEEE Journals}
%



\maketitle

\begin{abstract}
Accurate channel prediction is essential in massive multiple-input multiple-output (m-MIMO) systems to improve precoding effectiveness and reduce the overhead of channel state information (CSI) feedback. However, existing methods often suffer from accumulated prediction errors and poor generalization to dynamic wireless environments, making it challenging to maintain high prediction accuracy. Large language models (LLMs) have demonstrated remarkable modeling and generalization capabilities in tasks such as time series prediction, making them a promising solution. Nevertheless, a significant modality gap exists between the linguistic knowledge embedded in pretrained LLMs and the intrinsic characteristics of CSI, posing substantial challenges for their direct application to channel prediction. Moreover, the large parameter size of LLMs hinders their practical deployment in real-world communication systems with stringent latency constraints. To address these challenges, we propose a novel channel prediction framework based on semantically aligned large models, referred to as CSI-ALM, which bridges the modality gap between natural language and channel information. Specifically, we design a cross-modal fusion module that aligns CSI representations with the language feature space using a pretrained corpus. Additionally, we maximize the cosine similarity between word embeddings and CSI embeddings to construct semantic cues, effectively leveraging the latent knowledge in LLMs. To reduce complexity and enable practical implementation, we further introduce a lightweight version of the proposed approach, called CSI-ALM-Light. This variant is derived via a knowledge distillation strategy based on attention matrices, which extracts essential features from the teacher model, CSI-ALM, and transfers them to a compact, efficient student model, CSI-ALM-Light. Extensive experimental results demonstrate that CSI-ALM consistently outperforms state-of-the-art deep learning methods across various communication scenarios, achieving substantial performance gains. Moreover, under limited training data conditions—where all models are trained using only 10\% of the original training dataset—CSI-ALM-Light, with only 0.34M parameters, attains performance comparable to CSI-ALM and significantly outperforms conventional deep learning approaches. These validate the effectiveness  of the proposed approach for accurate and efficient channel prediction in m-MIMO systems.
\end{abstract}
\begin{IEEEkeywords}
Channel Prediction, Large Language Models, Massive MIMO, Knowledge Distillation, Modality Alignment.
\end{IEEEkeywords}

%
\IEEEpeerreviewmaketitle

\section{Introduction}
%
%
%
%
\IEEEPARstart{I}{N} modern wireless communication systems, massive multiple-input multiple-output (m-MIMO) technology has become a cornerstone for enhancing spectral efficiency \cite{b1,b2}. By deploying a large number of antennas at both the transmitter and receiver, these systems leverage spatial diversity to significantly outperform conventional systems in terms of throughput and robustness. However, the performance of MIMO systems heavily depends on the effectiveness of precoding techniques, which, in turn, require accurate and real-time channel state information (CSI)\cite{b3}. In frequency division duplex (FDD) systems, CSI feedback is essential but often suffers from latency due to feedback delays. In time division duplex (TDD) systems, although feedback requirements can be reduced by exploiting channel reciprocity between uplink and downlink, challenges remain in processing and updating CSI due to the time-varying nature of the channel\cite{b4,b5,b6}. Furthermore, in high-mobility scenarios, where the user’s position and velocity change rapidly, the accuracy of CSI estimation deteriorates, leading to significant performance degradation. To mitigate the adverse effects of channel aging and outdated CSI, extensive research has focused on channel prediction, which leverages the temporal correlation between historical and future channel states. Effective channel prediction techniques enable the system to anticipate future CSI, thereby improving precoding performance and enhancing overall system efficiency\cite{b7,b8,b9,b10,b11}.

Traditional methods for channel prediction are typically based on physical models or statistical learning approaches \cite{b12}. Early prediction methods relied heavily on modeling the physical characteristics of the wireless channel, such as multipath propagation, fading behavior, and noise. Based on these models, classical approaches such as MMSE-based channel estimation, Kalman filtering, and autoregressive (AR) models have been widely used for channel prediction \cite{b13,b14,b15}. While these methods can deliver accurate predictions under idealized conditions, their performance often degrades in real-world scenarios due to the oversimplified assumptions inherent in the models, which fail to capture the complex, nonlinear, and time-varying nature of wireless channels. Furthermore, as the scale of MIMO systems continues to grow, it becomes increasingly impractical to rely on model-based approaches for accurate channel modeling and prediction.

With the rapid advancement of deep learning (DL) techniques, data-driven methods have emerged as a promising solution to overcome the limitations of traditional model-based approaches. By leveraging large volumes of channel data, DL-based methods eliminate the need for complex mathematical modeling, enabling them to learn intricate, nonlinear patterns inherent in wireless channels \cite{b21,b22}. These approaches have been successfully applied using various architectures, including Convolutional Neural Networks (CNNs) \cite{b23}, Recurrent Neural Networks (RNNs) \cite{b24}, and their variants such as Long Short-Term Memory (LSTM) networks \cite{b25}, Gated Recurrent Units (GRUs) \cite{b26}, and Transformers \cite{b27}. While these models have demonstrated superior performance in channel prediction tasks, they often suffer from poor generalization when applied to new or unseen scenarios. Moreover, existing models typically require large amounts of training data, posing significant challenges in resource-constrained environments. Hence, improving model efficiency and adaptability while maintaining high prediction accuracy remains a critical challenge in channel prediction research.


The recent rise of Transformer-based architectures and large language models (LLMs) \cite{b30} has created new opportunities for channel prediction in wireless communications. Although originally designed for natural language tasks, LLMs possess a powerful ability to model long-range dependencies and sequential data, making them well-suited for capturing the temporal dynamics of wireless channels with improved accuracy and generalization. This cross-domain potential has led to several promising applications. For example, \cite{b31} introduces an LLM-based beam prediction method using textual prompts, achieving superior robustness and performance compared to traditional LSTM models. In \cite{b32}, LLMs are used for resource allocation, where fine-tuning on a small dataset delivers results comparable to advanced reinforcement learning techniques. Additionally, \cite{b33} explores a distributed training framework combining federated learning and aerial computing to support the deployment of large models in communication systems. \cite{b34} and \cite{b35} investigate the construction of efficient semantic communication systems using multimodal large models. Unlike conventional distortion-free transmission schemes, their approaches leverage large models to achieve high semantic fidelity while transmitting only minimal information. \cite{b36} further proposes a framework that integrates LLMs with multi-agent systems, enabling knowledge retrieval, task decomposition, and feedback mechanisms in 6G communications, thereby enhancing the intelligence and adaptability of communication systems.




Recent studies have begun exploring the application of pre-trained LLMs in channel prediction. For example, \cite{b37} proposes an LLM-based method for downlink channel prediction, where specialized embedding layers are designed for the frequency and angular domains while keeping the LLM backbone fixed. In \cite{b38}, the downlink CSI prediction task is reformulated to align with the autoregressive “next-token” generation mechanism of LLMs, resulting in a CSI-LLM framework that supports variable-length inputs and multi-step continuous prediction. However, these approaches do not fully address the fundamental differences between linguistic and channel data modalities. Directly applying pre-trained LLMs to channel prediction tasks may underutilize their representational capabilities, limiting the effective exploitation of the embedded knowledge within the models. Furthermore, the large parameter sizes and high computational complexity of LLMs present challenges for practical deployment in wireless communication systems.


To address these challenges, this paper proposes a novel channel prediction framework based on a pre-trained large language model (LLM), referred to as CSI-ALM, which leverages cross-modal semantic alignment and semantic prompting. Specifically, we design a customized embedding module tailored to capture meaningful representations from channel state information (CSI). To bridge the gap between channel and linguistic modalities, we introduce a cross-modal semantic alignment module that aligns CSI embeddings with pre-trained word embeddings. Additionally, to enhance the aligned features, we incorporate the most semantically relevant word embeddings, selected based on cosine similarity. To reduce prediction latency and enable practical deployment, we also propose a lightweight variant of CSI-ALM, termed CSI-ALM-Light. This version is obtained through a knowledge distillation approach that utilizes self-attention matrices to extract essential knowledge from the teacher model, CSI-ALM, and transfer it to a computationally efficient student model, CSI-ALM-Light. Experimental results validate the effectiveness of the proposed approach in terms of prediction accuracy, as well as the efficiency and robustness of the lightweight version, which achieves competitive performance with significantly fewer parameters under limited data conditions. The main contributions of this paper can be summarized as follows:

\begin{itemize}
    \item We propose a novel channel prediction scheme, called CSI-ALM, along with its lightweight variant, CSI-ALM-Light, based on a pre-trained large language model, leveraging modality alignment and knowledge distillation to achieve both high prediction accuracy and computational efficiency. The framework addresses the knowledge misalignment between the linguistic representations in the pre-trained LLM and the channel data, while also offering a practical solution for lightweight LLM deployment.
    
    \item To bridge the modality gap between CSI data and linguistic data, we design a dedicated embedding layer for channel data to obtain CSI embeddings. Additionally, we design a cross-modal attention mechanism, integrated with the pre-trained corpus of LLMs, to align the CSI embeddings with the text feature space of the LLM. We also utilize maximum cosine similarity-based retrieval to obtain semantic prompts for the CSI, which are used as prefixes to further achieve semantic space alignment.
    
    \item To obtain the lightweight model, we adopt a knowledge distillation approach that leverages the self-attention mechanisms of Transformer-based models. Specifically, the distillation process transfers CSI knowledge from the teacher model to the student model by minimizing the Kullback–Leibler (KL) divergence between their respective self-attention distributions.
    
    \item Extensive experimental results demonstrate that CSI-ALM consistently outperforms state-of-the-art deep learning methods across various communication scenarios, achieving substantial performance gains. Furthermore, in scenarios with limited training data—using just 10\% of the original dataset—CSI-ALM-Light, with only 0.34 million parameters, achieves accuracy comparable to CSI-ALM while substantially outperforming other baseline approaches. These results highlight the proposed framework’s effectiveness in enabling both accurate and efficient channel prediction for m-MIMO systems.

\end{itemize}

The rest of this article is organised as follows. Section II introduces the MIMO channel prediction model and performance metrics studied. Section III provides a detailed description of the proposed CSI-ALM framework and the modality alignment method. Section IV introduces the proposed CSI-ALM-Light and the self-attention relationship-based knowledge distillation method. Section V introduces the model training method. Section VI discusses the simulation results, and Section VII concludes the paper.

\textbf{Notation:} Bold letters denote vectors and matrices, while scalars are represented by plain letters. For a vector $\bm{x}$, the $i$th component is written as $x_i$. For a matrix $\bm{Y}$, $\bm{Y} \in \mathbb{R}^{M \times N}$ specifies that $\bm{Y}$ is of dimensions $M \times N$. The notation $\mathcal{CN}(\mu, \sigma^2)$ represents a multivariate circular complex Gaussian distribution with mean $\mu$ and  $\sigma^2$.

\section{System model}
\subsection{Channel Model}

For simplicity but without loss of generality, We consider a massive MIMO system where the base station (BS) is equipped with $M \gg 1$ antennas arranged in a uniform linear array (ULA), serving multiple single-antenna users randomly distributed in $B$ regions. Users within the same region are assumed to share a common propagation environment. The channel state information $\bm{\mathit{H}}(f)$ between a user and the BS at carrier frequency $f$ can be modeled as a combination of multipath components:

\begin{equation}
    \bm{\mathit{H}}(f) = \int_{\theta \in \mathcal{A}_k} \alpha(\theta) \mathbf{a}(\theta) e^{-j 2 \pi f \tau(\theta) + j \phi(\theta)} d\theta,
\end{equation}where $\alpha(\theta)$ is the attenuation coefficient of the signal path from direction $\theta$, $\phi(\theta)$ is the phase shift, $\tau(\theta)$ is the delay associated with the direction of arrival (DOA) $\theta$, and $\mathbf{a}(\theta) \in \mathbb{C}^{M \times 1}$ is the array response vector defined as:
    \begin{equation}
        \mathbf{a}(\theta) = \left[1, e^{-j 2 \pi d \sin(\theta) / \lambda}, \cdots, e^{-j 2 \pi (M-1) d \sin(\theta) / \lambda}\right]^T,
    \end{equation}
where $d$ is the antenna spacing, and $\lambda$ is the signal wavelength.

We note that the uplink channel $\bm{\mathit{H}}_u$ and downlink channel $\bm{\mathit{H}}_d$ exhibit similar propagation characteristics. A deterministic mapping exists between them, denoted by $\Psi: \bm{\mathit{H}}_u \mapsto \bm{\mathit{H}}_d$. This mapping, though complex and nonlinear, can be learned by a model—such as a transformer or a large pre-trained neural network—trained on a sufficiently large dataset.

\subsection{Downlink CSI Prediction}

Downlink CSI prediction involves using uplink CSI data from previous transmission time steps to predict the downlink CSI at the next time step. The functional mapping between the downlink CSI at time $t$ and the uplink CSI sequence from the past \( T \) steps can be given by:

\begin{equation}
    \bm{{\mathit{H}}}_d^t = h({\bm{\mathit{H}}_u^{t-1}}, {\bm{\mathit{H}}_u^{t-2}}, \dots, {\bm{\mathit{H}}_u^{t-T}}).
\end{equation}

Our goal is to learn a function $g(\cdot)$ that approximates the true mapping $h(\cdot)$,  by minimizing the normalized mean squared error (NMSE) between the predicted and true downlink CSI:
\begin{equation}
    \min \mathcal{L}_{NMSE} = \mathbb{E} \left\{\frac{\| \bm{{\mathit{H}}}_d^t - \bm{\hat{\mathit{H}}}_d^t \|^2}{\| \bm{{\mathit{H}}}_d^t \|^2}\right\},
\end{equation}
\begin{equation}
    \text{s.t.}    \bm{\hat{\mathit{H}}}_d^t = g({\bm{\mathit{H}}_u^{t-1}}, {\bm{\mathit{H}}_u^{t-2}}, \dots, {\bm{\mathit{H}}_u^{t-T}}),
\end{equation}
where \(\bm{{\mathit{H}}}_d^t\) denotes the true downlink CSI at time $t$, and \(\bm{\hat{\mathit{H}}_d}^t\) represents the predicted downlink CSI. 

\subsection{ Performance Metrics }

NMSE is a widely used evaluation metric in tasks such as channel prediction and signal processing. It is used to measure the magnitude of the error between predicted values and actual values, particularly in the context of channel prediction, where NMSE quantifies the accuracy of CSI. To evaluate the accuracy of the CSI prediction, we use NMSE, which is calculated by:

\begin{equation}
    \text{NMSE} = \frac{\| \bm{\mathit{H}} - \bm{\hat{\mathit{H}}} \|^2}{\| \bm{\mathit{H}} \|^2},
\end{equation}
where \(\bm{\mathit{H}}\) is the true CSI and \(\hat{\bm{\mathit{H}}}\) is the predicted CSI. A lower NMSE value indicates more accurate CSI prediction.

\begin{figure*}[htb]
  \centering
  \includegraphics[width=\textwidth]{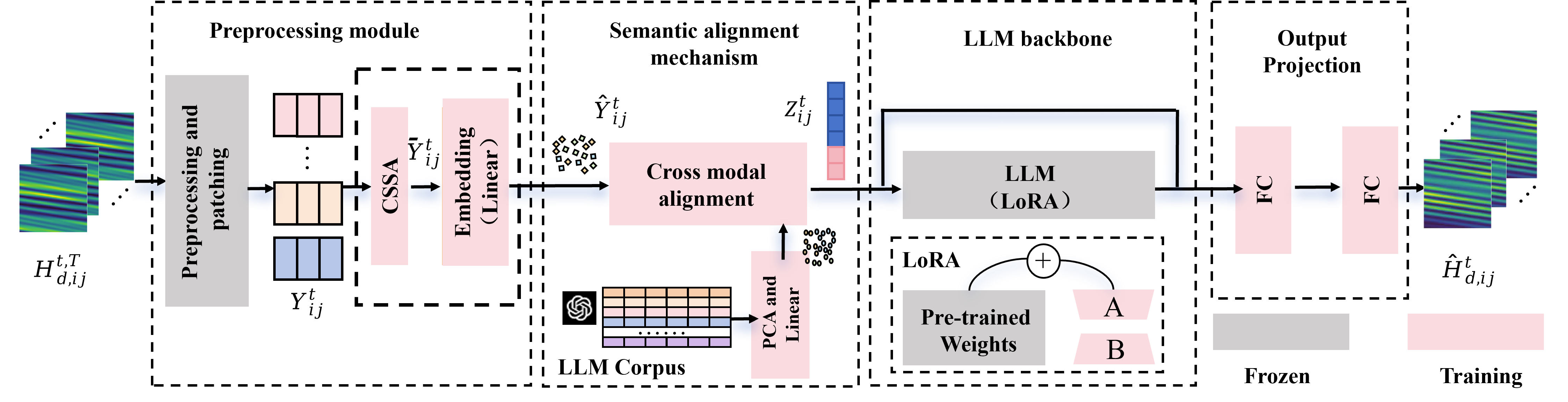}
  \caption{Our proposed CSI-ALM.}
  \label{fig.1}
\end{figure*}
\section{Proposed CSI-ALM system}
In this section, we present our proposed LLM-based pipeline for channel prediction, as well as the modality alignment mechanism that aligns the semantics of CSI and natural language. 
\subsection{LLM based Pipeline}
The LLM-based pipeline for channel prediction is illustrated in Fig.~\ref{fig.1}. It consists of a preprocessing module for CSI data with a semantic alignment mechanism, a fine-tuned LLM backbone, and a final output projection module. The CSI data preprocessing module is used to map the original CSI data to the feature space, and then align the CSI features to the text semantic space of LLM through a semantic alignment mechanism. Finally, after sequence modeling by LLM and adjusting the output dimension through the final output layer, the predicted CSI is obtained. We introduce each module in detail in the following.

The input uplink CSI sequence first undergoes a series of preprocessing steps. However, directly feeding the high-dimensional sequence into the network can lead to substantial computational complexity and prolonged training times, particularly when the number of antennas and subcarriers is large. To address this efficiently, we propose a parallel antenna processing approach, where the CSI for each transmitter–receiver antenna pair $(i, j)$is predicted separately. Here, $i$ and $j$ denote the indices of the transmitter and receiver antennas, respectively. Specifically, the input sequence at time $t$ for each pair $\bm{\mathit{H}}^{t,T}_{u, ij} \in \mathbb{C}^{F \times T}$ is the CSI sequence between the $i$-th transmitting antenna and the 
$j$-th receiving antenna over the past $T$ time steps:
\begin{equation}
        \bm{\mathit{H}}^{t,T}_{u, ij} = [ \bm{\mathit{H}}^{t-1}_{u, ij}, ...,  \bm{\mathit{H}}^{t-T}_{u, ij}],
\end{equation}
where $ \bm{\mathit{H}}^{t-p}_{u, ij}$ is the CSI of the transmitter antenna $i$ and the receiver $j$ at time $t-p$, $F$ denotes the number of subcarriers per antenna pair. We recall that $T$ is the temporal length of historical CSI used for prediction.

    
Since neural networks typically operate on real numbers, we convert $\bm{\mathit{H}}^{t,T}_{u, ij}$ into real tensors $\bm{\mathit{X}}^{t}_{ ij}\in \mathbb{R}^{2 \times F \times T}$. To facilitate network training and improve convergence, we further normalize the input data as follows:
\begin{equation}
        \tilde{\bm{\mathit{X}}}^{t}_{ ij} = \frac{\bm{\mathit{X}}^{t}_{ ij} - \mu}{\sigma}, \quad
\end{equation}
where $\mu$ and $\sigma $ denote the mean and standard deviation of the real-valued sequence within a training batch.

Then, $\tilde{\bm{\mathit{X}}}^{t}_{ij}$ is reshaped into a tensor of dimension $2F \times T$. To capture local temporal features and reduce computational complexity, we apply a patching operation along the temporal dimension, as illustrated in Fig.\ref{fig.1}. Specifically, $\tilde{\bm{\mathit{X}}}^{t}_{ij}$ is divided into non-overlapping patches of size $N$, resulting in a tensor $\bm{\mathit{Y}}^{t}_{ij}\in \mathbb{R}^{2F \times N \times T'}$,
where $T' = \left\lceil \frac{T}{N} \right\rceil$ denotes the number of patches. Zero-padding is applied to the final patch if it is not completely filled.

\begin{figure}
        \centering
        \includegraphics[width=1\linewidth]{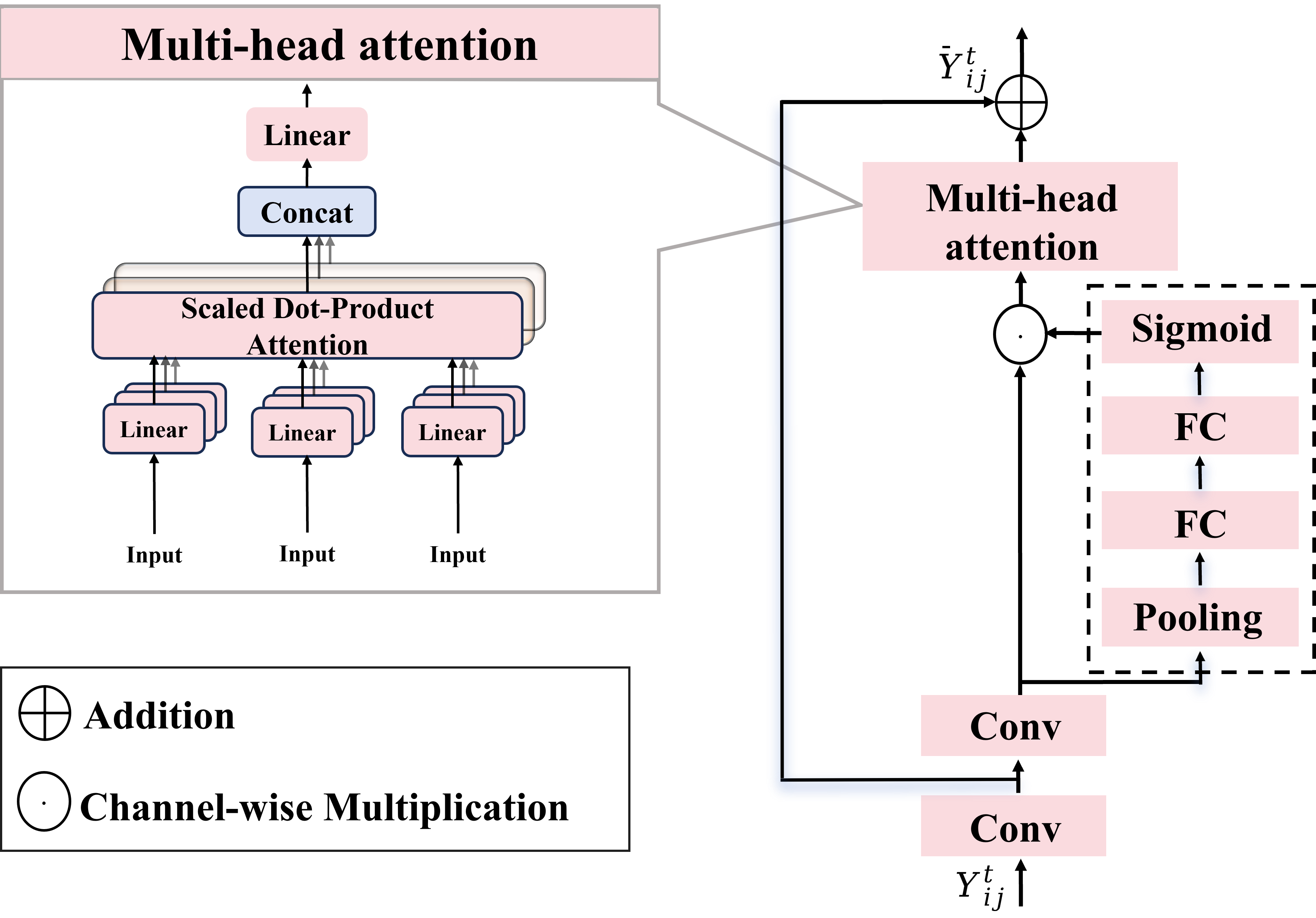}
        \caption{The Cross Spatial Self-Attention (CSSA) module.}
        \label{fig.2}
    \end{figure}

After serializing the CSI via the patching operation, we introduce a Cross Spatial Self-Attention (CSSA) module, as illustrated in Fig.~\ref{fig.2}. CSSA improves the modeling ability of the model for key channel features by dynamically adjusting the importance of focusing on different channel characteristics, effectively capturing inter channel interference patterns:
\begin{equation}
    \overline{\bm{\mathit{Y}}}^{t}_{ij} = \text{Reshape}(\text{CSSA}\left( \bm{\mathit{Y}}^{t}_{ij} \right)),
\end{equation}where $\overline{\bm{\mathit{Y}}}^{t}_{ij} \in \mathbb{R}^{2F \times T}$represents the latent CSI features that have been modulated and reshaped by the CSSA module.

To better capture the underlying characteristics of the channel and adapt to the input format required by the subsequent LLM backbone, we employ an embedding layer to project the original channel information into a high-dimensional feature space $\hat{\bm{\mathit{Y}}}^{t}_{ij} \in \mathbb{R}^{D \times T}$: 
\begin{equation}
\hat{\bm{\mathit{Y}}}^{t}_{ij}  = \text{Linear}(\overline{\bm{\mathit{Y}}}^{t}_{ij}), \quad
\end{equation}
Next, through a cross-modal attention alignment mechanism, $\hat{\bm{\mathit{Y}}}^{t}_{ij}$  is transformed into corresponding text $\hat{\bm{\mathit{Y}}}^{t}_{ij, \mathrm{text}}$:
\begin{equation}
    \hat{\bm{\mathit{Y}}}^{t}_{ij, \mathrm{text}} = \text{Crossmodal-Alignment}(\hat{\bm{\mathit{Y}}}^{t}_{ij}),
\end{equation}
The details of the cross-modal attention alignment process are provided in Section III.B.

After the cross-modal alignment process, the textual CSI representation $\hat{\bm{\mathit{Y}}}^{t}_{ij, \mathrm{text}}$ is combined with a semantic prompt $\bm {e}' \in \mathbb {R}^{D\times l}$ to obtain $\bm{Z}^{t}_{ij}$, which is then input into the LLM backbone network for further processing. The procedure for generating semantic prompts is described in Section III.B. In this work, we adopt GPT-2 as the LLM backbone and fine-tune it using LoRA \cite{b39} during training. The LLM maps the refined latent CSI sequence to a prediction of the current CSI feature, which is subsequently passed through a projection layer to produce the final CSI prediction:
\begin{equation}
        \bm{\hat{\mathit{H}}}_{d, ij}^t = \text{Proj}\left( \text{LLM} (\bm{Z}^{t}_{ij}) ) \right),
\end{equation}where $\bm{\hat{\mathit{H}}}_{d, ij}^t\in \mathbb{R}^{2F} $  denotes the final predicted downlink CSI at time $t$. Through this parallel processing method, we can obtain the downlink channel CSI $\bm{\hat{\mathit{H}}}_d ^{t}$ of the MIMO system.

 \subsection{Modality Alignment Mechanism}
\begin{figure}
        \centering
        \includegraphics[width=1\linewidth]{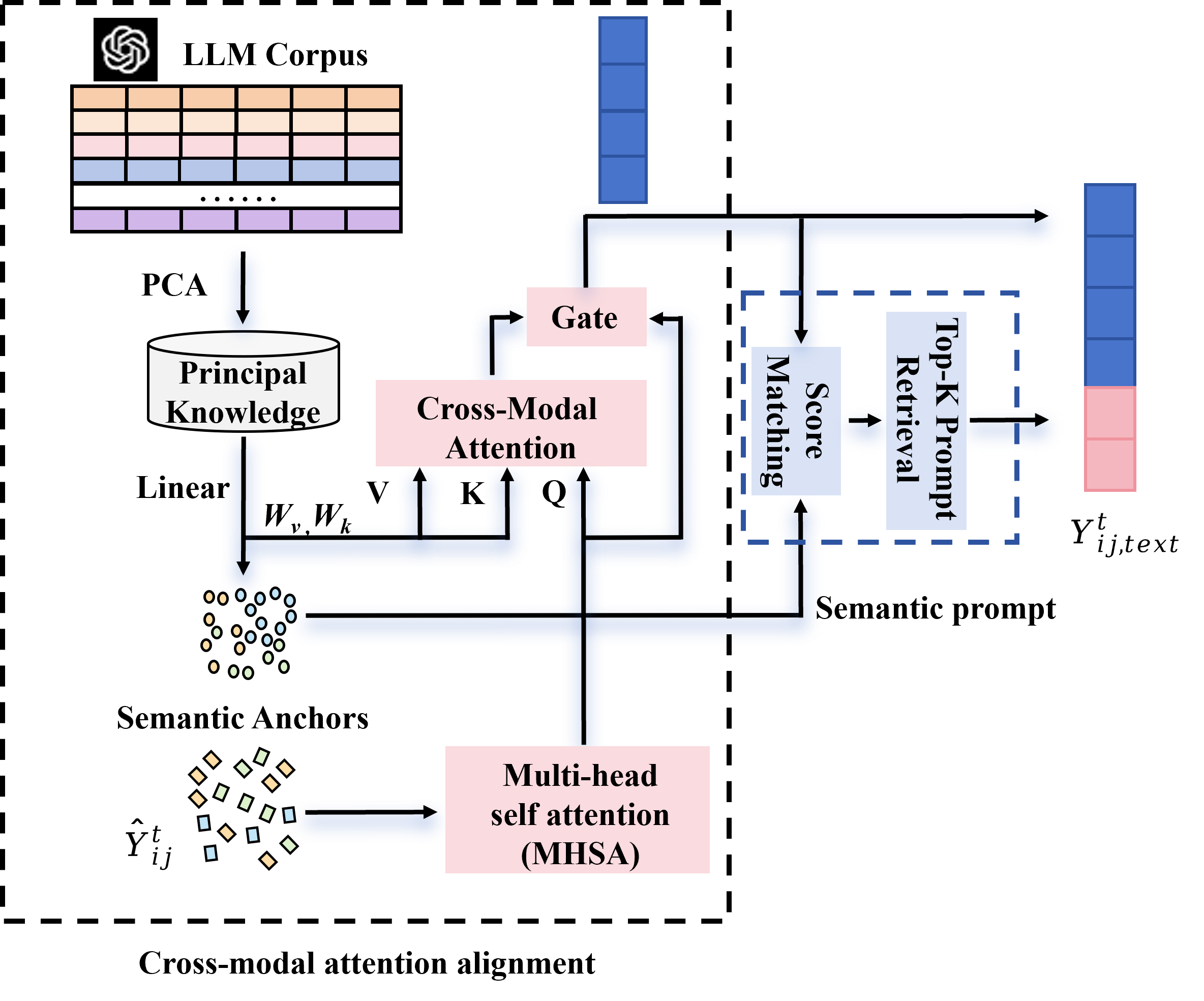}
        \caption{The modality alignment mechanism.}
        \label{fig.5}
    \end{figure}

To enable modality alignment between channel information and the linguistic knowledge embedded in a pre-trained LLM, we propose a modality alignment method that leverages the LLM's corpus, as illustrated in Fig.~\ref{fig.5}. This method consists of two key components: cross-modal attention-based feature alignment module and semantic prompt module.
\begin{enumerate}[label=\arabic*), wide, labelwidth=0pt, labelsep=0.5em, align=left]
\item 
\textbf{Cross-Modal Attention Feature Alignment Module:} This module transfers language knowledge from a pre-trained LLM to CSI prediction, achieving cross-modal feature alignment to leverage the LLM's pre-trained capabilities. Specifically, we first derive the projected time sequence from the latent channel features, output by the embedding module, using a multi-head self-attention (MHSA) mechanism:
\begin{equation}
    {\bm{\mathit{Y}}}^{t}_{ij,\mathrm{time}} = \text{MHSA}(\hat{\bm{\mathit{Y}}}^{t}_{ij}).
\end{equation}

We then apply a cross-attention module to align ${\bm{\mathit{Y}}}^{t}_{ij,time}$ with the word embedding dictionary $\mathcal{D} \in \mathbb{R}^{|A| \times D}$ of the pretrained LLM, where $|A|$ is the vocabulary size and and $M$ is the embedding length. However, $|A|$ is typically very large, e.g., 50,257 for GPT-2, making direct cross-attention alignment computationally prohibitive. To address this, we apply Principal Component Analysis (PCA) along the vocabulary dimension of $\mathcal{D}$ to reduce its dimensionality. This yields a compact dictionary of linearly combined embeddings instead of the full set, significantly reducing the number of entries:
\begin{equation}
\hat{\mathcal{D}} = \text{PCA}(\mathcal{D}), d \ll |A|.    
\end{equation}
Here, $d$ is a predefined reduced dictionary size. Notably, this PCA step is performed only once before training, incurring negligible overhead during optimization.

In order to better match the pre trained text knowledge and map it to the text space, we use a pre trained vocabulary library to perform cross modal attention fusion on CSI features, where $\hat{\mathcal{D}}$ is used to generate the key and value matrices, and ${\bm{\mathit{Y}}}^{t}_{ij,\mathrm{time}}$ is used to generate the query matrix. This aligns the principal word embeddings with the temporal latent CSI representation to obtain the aligned textual representation ${\bm{\mathit{Y}}}^{t}_{ij,\mathrm{cross}} \in \mathbb{R}^{D \times T}$:
\begin{equation}
{\bm{\mathit{Y}}}^{t}_{ij,\mathrm{cross} }= \text{Softmax}\left(\frac{Q K^T}{\sqrt{c}}\right)V,    
\end{equation}where
\begin{equation}
Q = {\bm{\mathit{Y}}}^{t}_{ij,\mathrm{time}}W_Q, \quad K = \hat{\mathcal{D}}W_K, \quad V = \hat{\mathcal{D}}W_V,    
\end{equation}
and $W_Q, W_K, W_V \in \mathbb{R}^{D \times D}$ are the projection matrices for the query ($Q$), key ($K$), and value ($V$) matrices, respectively. The scalar $c$ is a scaling factor. The aligned textual representation and temporal representation are then adaptively combined to produce the final representation ${\bm{\mathit{Y}}}^{t}_{ij,\text{text}}$:
\begin{equation}
    g = \text{Sigmoid}\left(W_{\text{gate}}^T \left[{\bm{\mathit{Y}}}^{t}_{ij,\text{cross}} ; {\bm{\mathit{Y}}}^{t}_{ij,\text{time}} \right]\right),
\end{equation}
\begin{equation}
    {\bm{\mathit{Y}}}^{t}_{ij,\text{text}} = {\bm{\mathit{Y}}}^{t}_{ij,\text{cross}} \odot g + {\bm{\mathit{Y}}}^{t}_{ij,\text{time}}  \odot (1 - g),
\end{equation}
where $W_{\text{gate}} \in \mathbb{R}^{2T \times D}$ is the learned gating projection matrix, and $\odot$ denotes element-wise multiplication.
    \item 
\textbf{Semantic prompt module:}
This module leverages the pre-trained corpus $\mathcal{D}$ to generate semantic prompts that enhance the input to the LLM. Similarly, we transform the pre-trained word embedding dictionary into a compact set of semantic anchors $\hat{\mathcal{D}}'\in \mathbb{R}^{m \times D}$, where $m \textless d$. Unlike previous approaches that apply PCA, we use a learnable linear layer for this transformation:
\begin{equation}
    \hat{\mathcal{D'}} = \text{Linear}(\hat{\mathcal{D}}).    
\end{equation}

To enrich the CSI embeddings, we apply cosine similarity-based score matching to retrieve the most relevant semantic anchors:
    \begin{equation}
    \gamma \left( {\bm{\mathit{Y}}}^{t}_{ij,\text{text}}, \bm{\mathit{e}}' \right) =
    \frac{{\bm{\mathit{Y}}}^{t}_{ij,\text{text}} \cdot \bm{\mathit{e}}'}
    {\|{\bm{\mathit{Y}}}^{t}_{ij,\text{text}}\| \|\bm{\mathit{e}}'\|},
    \end{equation}
where $\bm{\mathit{e}}' \in \hat{\mathcal{D}}'$. We then select the top $K$ highest-scoring embedding vectors as semantic prompts $e'_{1:K}$. These selected prompts are concatenated as a prefix to ${\bm{\mathit{Y}}}^{t}_{ij,\text{text}}$ to form the enhanced representation ${\bm{\mathit{Z}}}^{t}_{ij}$, which is subsequently fed into the LLM backbone:
    \begin{equation}
        {\bm{\mathit{Z}}}^{t}_{ij} =
        \begin{bmatrix}
            e'_{1}, \dots, e'_{K}; {\bm{\mathit{Y}}}^{t}_{ij,\text{text}}
        \end{bmatrix}
    \end{equation}   
\end{enumerate}

\section{A Lightweight variant}
LLMs have a large number of parameters and experience long inference latency, making them unsuitable for the stringent requirements of edge deployment and real-time processing in wireless communication systems. In this section, we propose a knowledge distillation (KD) method that leverages self-attention relationships to transfer knowledge from CSI-ALM to its lightweight variant, CSI-ALM-Light.

\subsection{Knowledge Distillation}

    \begin{figure}
        \centering
        \includegraphics[width=0.9\linewidth]{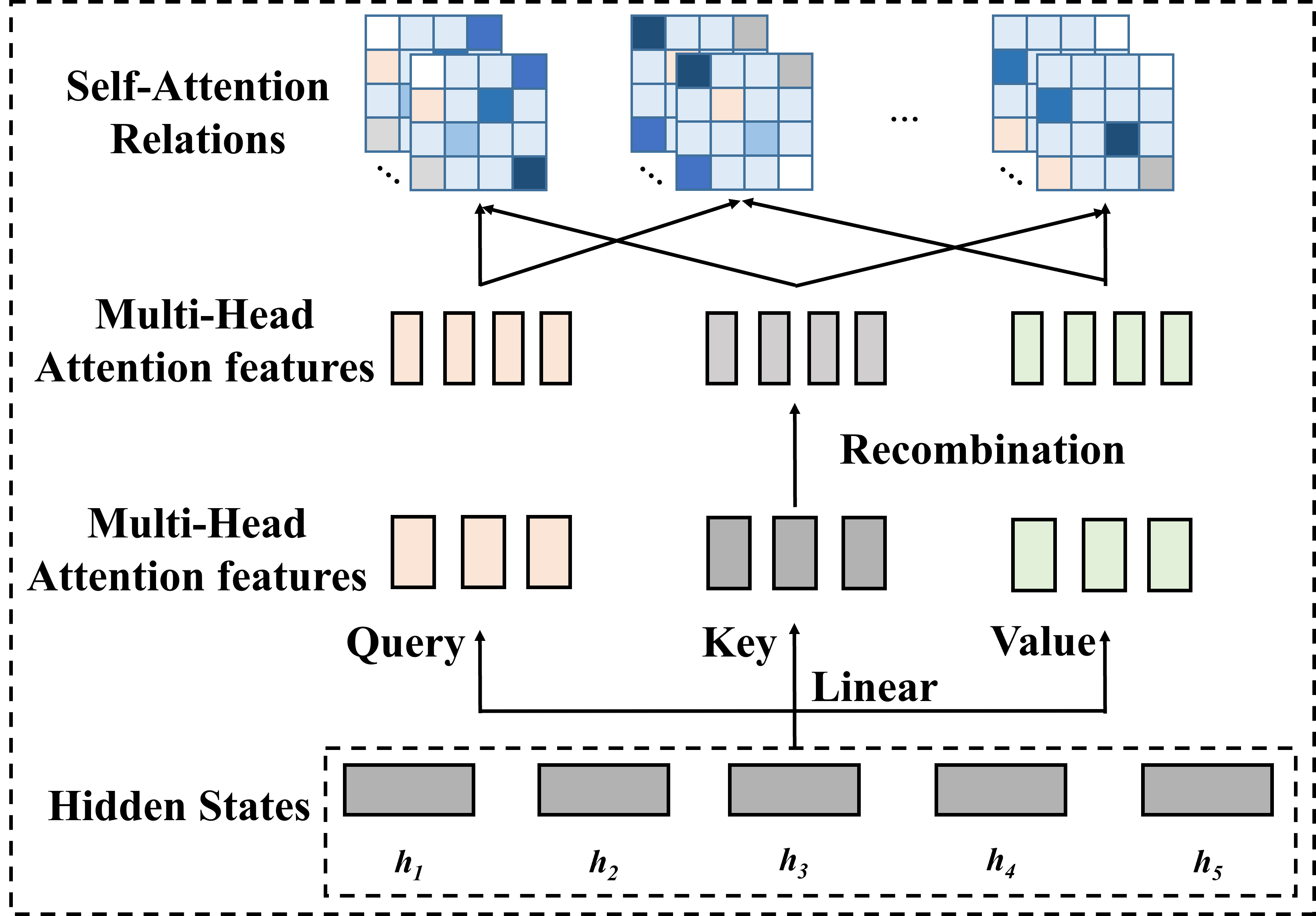}        \caption{Self-Attention Relations.}
        \label{fig3}
    \end{figure}
    
The core of our method lies in leveraging multi-head attention relations, derived from the scaled dot-product of queries, keys, and values across multiple attention heads, for knowledge distillation between the teacher and student models. As illustrated in Fig.$\ref{fig3}$, we first concatenate the queries, keys, and values from different attention heads, and then split the concatenated vector into individual components according to the desired number of attention heads. The same procedure is applied to the keys and values. This approach enables a more flexible handling of the self-attention mechanism, particularly in scenarios where the teacher and student models employ differing numbers of attention heads. After reorganizing the queries, keys, and values, we proceed to compute the attention relations of both models:
\begin{equation}
\bm{R}_{xy,a}^{\text{teacher}} = \text{softmax}\left(\frac{\bm{A}_x^{a,{\text{teacher}}} \bm{(A}_y^{a,{\text{teacher}})^T}}{\sqrt{d_r}}\right),
\end{equation}
\begin{equation}
\bm{R}_{xy,a}^{\text{student}} = \text{softmax}\left(\frac{\bm{A}_x^{a,{\text{student}}} \bm{(A}_y^{a,{\text{student}}})^T}{\sqrt{d_r'}}\right),
\end{equation}where $\bm{A}_x^{a,{\text{teacher}}} \in \mathbb{R}^{|n| \times d_r}$ and $\bm{A}_x^{a,{\text{student}}} \in \mathbb{R}^{|n| \times d_r'}$ ($x \in [1, 3]$) are the queries, keys and values of the relationship header between the last teacher layer and student layer. $d_r$ and $d_r'$ are the relation head size of teacher and student models. $\bm{R}_{xy,a}^{\text{teacher}} \in \mathbb{R}^{|n| \times |n|}$ is the self-attention relation of a teacher’s relation head and $\bm{R}_{xy,a}^{\text{student}} \in \mathbb{R}^{ |n| \times |n|}$ is the self-attention relation of a student model's relation head. 

\begin{figure}
        \centering
        \includegraphics[width=1\linewidth]{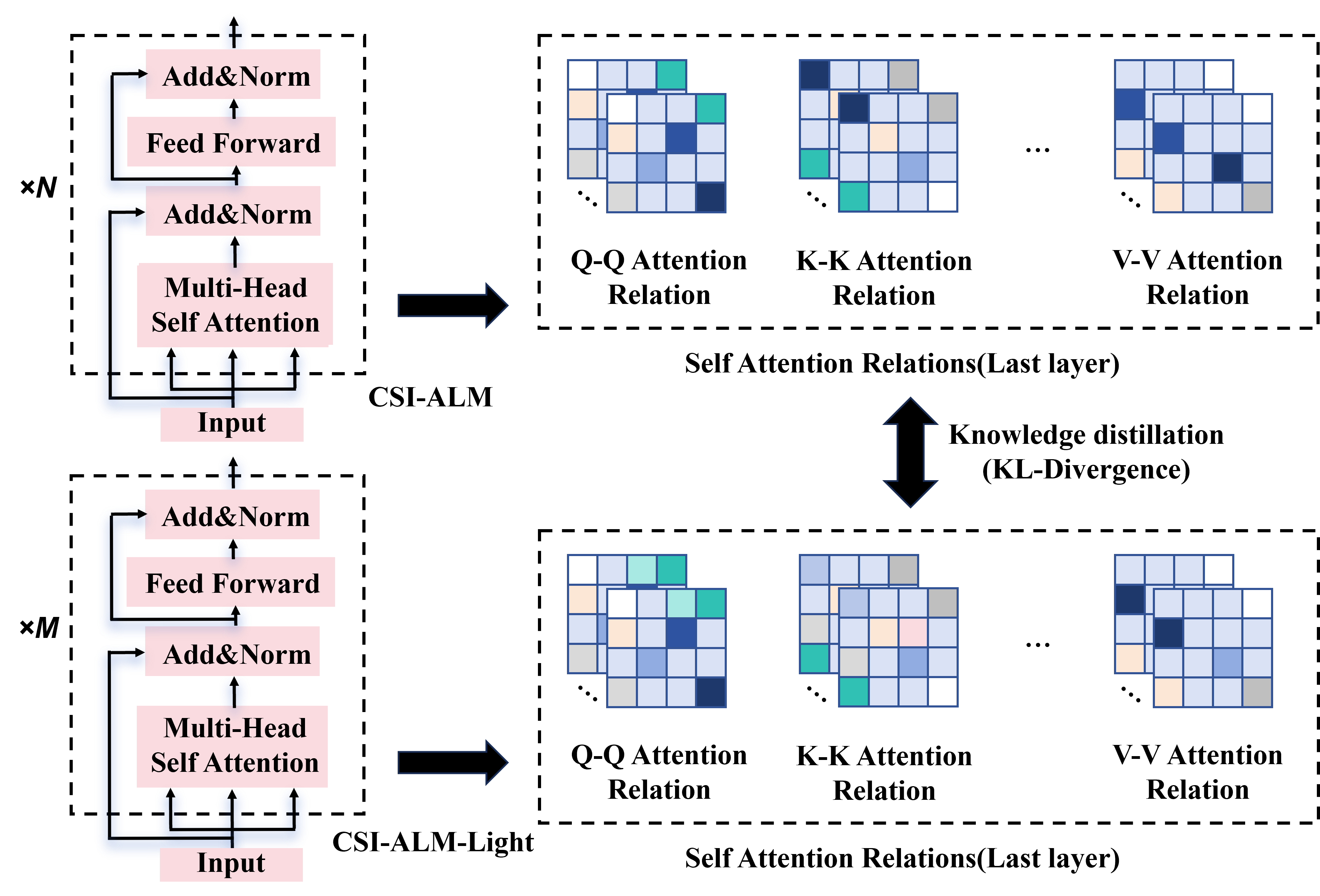}        \caption{Self attention Knowledge
Distillation.}
        \label{fig.4}
    \end{figure}

In traditional knowledge distillation approaches, the architecture of the teacher model often constrains the number of attention heads that the student model can employ. However, our method overcomes this limitation by aligning the queries, keys, and values from both the teacher and student models into a shared vector space with a uniform number of relational heads, regardless of their original attention head configurations. This alignment strategy enables the student model to effectively learn from the teacher model even when there is a discrepancy in their attention head configurations. As illustrated in Fig.$\ref{fig.4}$, after obtaining the attention relationship between the teacher model and the student model, we use the KL divergence between the attention relationships of the teacher model and the student model as the goal of knowledge distillation:
\begin{equation}
    \mathcal{L} = \sum_{x=1}^{3} \sum_{y=1}^{3} \alpha_{xy} \mathcal{L}_{xy},
\end{equation}
\begin{equation}
\mathcal{L}_{xy} = \frac{1}{|A_r|} \sum_{a=1}^{A_r}  D_{\text{KL}}\left( \bm{R}_{xy,a}^{\text{teacher}} \| \bm{R}_{xy,a}^{\text{student}} \right),
\end{equation}where $A_r$ is the number of relation heads. If the number of relation heads and attention heads is the same, the Q-K relation is equivalent to the attention weights in the self-attention module. $\alpha_{xy} \in \{0, 1\}$ is the weight assigned to each self-attention relation loss. We transfer query-query, key-key and value-value self-attention relations to balance the performance and training cost.
    
By using more relation heads when calculating self attention relationships, our method provides more fine-grained self attention knowledge for student models. This enhanced self attention learning helps student models capture more detailed relationships between input features, thereby improving performance and deployment efficiency. Our method effectively transfers rich attention knowledge from the teacher model to the student model, ensuring that the student model performs well while reducing complexity, making it suitable for resource constrained environments such as edge devices.

\subsection{CSI-ALM-Light}
\begin{figure}
        \centering
        \includegraphics[width=0.6\linewidth]{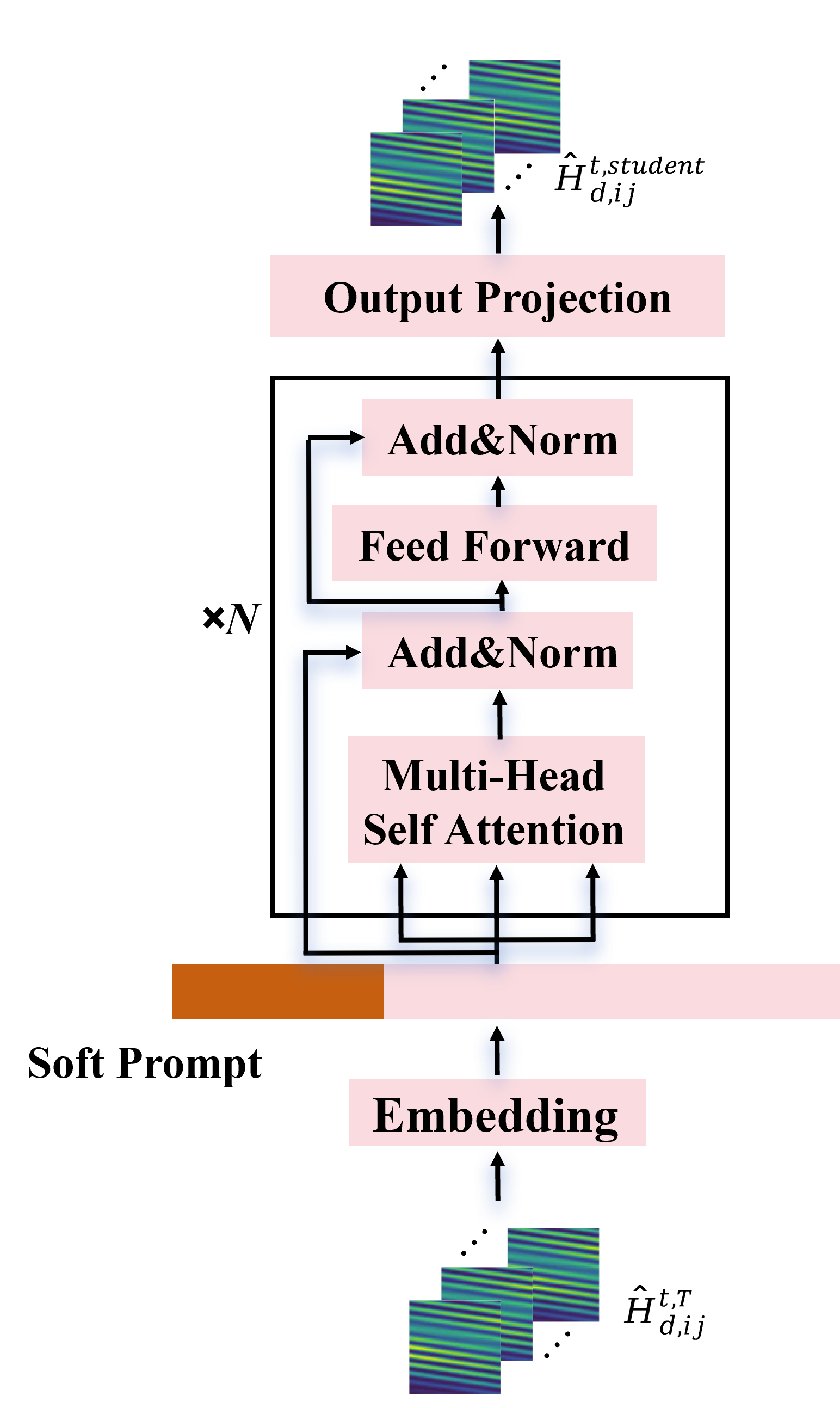}
        \caption{CSI-ALM-Light.}
        \label{fig4}
    \end{figure}

In order to meet the requirements of inference speed and computational complexity in practical deployment, we introduce a small yet sophisticated student model, called CSI-ALM-Light. As shown in Fig.$\ref{fig4}$, CSI-ALM-Light adopts the same decoder-only architecture as GPT-2. Compared with the more complex structure of CSI-ALM, CSI-ALM-Light uses fewer hidden state dimensions and layers, retaining only a simple CSI embedding module and the final projection layer. Similar to CSI-ALM, we also use a parallel approach to process antennas here, inputting $\bm{\mathit{H}}^{t,T}_{u, ij} $passes through the embedding layer to obtain CSI embedding:
\begin{equation}
    \bm{\mathit{Y}}_{ij}^{t,\text{student}} = \text{Embedding}(\bm{\mathit{H}}^{t,T}_{u, ij})
\end{equation}where $\bm{\mathit{Y}}_{ij}^{t, \text{student}}$ is the CSI embedding of the student model, which is further fed into the transformer layer.

In order to further improve the learning ability of the CSI-ALM-Light, we introduced a set of learnable parameters of a certain length as soft prompts for the CSI-ALM-Light, connected them with CSI embedding $\bm{\mathit{Y}}_{ij}^{t,\text{student}}$, and sent them to the transformer layer for processing:
    \begin{equation}
        \hat{\bm{\mathit{Y}}}_{ij}^{t,\text{student}} = \text{transformer} (\text{Concat}  (\bm{e}^{\text{student}}, \bm{\mathit{Y}}_{ij}^{t,\text{student}}) ), 
    \end{equation}where $\bm{e}^{\text{student}}$ are soft prompts and $\hat{\bm{\mathit{Y}}}_{ij}^{t,\text{student}}$ is output features of transformer layer.
    
Finally, the hidden state is processed through the output projection layer to obtain the CSI predicted by the CSI-ALM-Light:
\begin{equation}
     \bm{\hat{\mathit{H}}}_{d, ij}^{t,\text{student}}= \text{Proj}(\hat{\bm{\mathit{Y}}}_{ij}^{t,\text{student}}),
\end{equation}where $\bm{\hat{\mathit{H}}}_{d, ij}^{t,\text{student}}$ is the CSI between the $i$-th transmitting antenna and the $j$-th receiving antenna at time $t$. Through this parallel processing method, we can obtain the downlink channel CSI $\bm{\hat{\mathit{H}}}_d ^{t,\text{student}}$ of the MIMO system.

\section{MODEL TRAINING}
In this section, we introduce the training methods for our CSI-ALM and the lightweight knowledge distillation approach. In the knowledge distillation training of the proposed CSI-ALM and the student model, we first train the original CSI-ALM. During this process, we simultaneously consider the impact of the task objectives on model training and the embedding alignment loss of semantic prompts:
\begin{equation}
\mathcal{L}_{all} =  \mathcal{L}_{NMSE} -  \lambda_1 \mathcal{L}_{ce},
\end{equation}where $\mathcal{L}_{NMSE}$ represents the NMSE loss between the predicted CSI and the ground truth CSI, which is given by :
\begin{equation}
        \mathcal{L}_{NMSE}(\bm{\hat{\mathit{H}}}_d^t, \bm{{\mathit{H}}}_d^t) = \mathbb{E} \left\{\frac{\| \bm{{\mathit{H}}}_d^t - \bm{\hat{\mathit{H}}}_d^t \|^2}{\| \bm{{\mathit{H}}}_d^t \|^2}\right\},
\end{equation}where $\bm{H}_{d}$ and $\bm{\hat{\mathit{H}}_d} $ represent real CSI and predicted CSI, respectively.
While $\mathcal{L}_{ce}$ the alignment loss between the top-k selected semantic prompts and the CSI embeddings, which is given by:
\begin{equation}
    \mathcal{L}_{ce}({\bm{\mathit{Y}}}^{t}_{ij,\text{text}}, e'_{top-K})=    \frac{{\bm{\mathit{Y}}}^{t}_{ij,\text{text}} \cdot \bm{\mathit{e'_{top-K}}}}
    {\|{\bm{\mathit{Y}}}^{t}_{ij,\text{text}}\| \|\bm{\mathit{e'_{top-K}}}\|},
\end{equation}where ${\bm{\mathit{Y}}}^{t}_{ij,\text{text}}$ represents the CSI features after cross modal attention alignment, and $e'_{top-K}$ represents the semantic prompts after similarity matching. This loss aims to maximize the similarity between the semantic prompts and the CSI embedding feature space to achieve alignment in the semantic space.
\begin{algorithm}
\caption{Training Algorithm of the CSI-ALM}
\label{alg:MACSII-LLM}
\SetKwInOut{Input}{Input}
\SetKwInOut{Output}{Output}

\Input{Historical channel information ${\bm{\mathit{H}}_u^{t-1}}, {\bm{\mathit{H}}_u^{t-2}}, \dots, {\bm{\mathit{H}}_u^{t-T}}$ and true values of predicted channel information $\bm{{\mathit{H}}}_d^t$}
\While{Stop criterion is not met}{
    Output cross modal features ${\bm{\mathit{Y}}}^{t}_{ij,\text{text}}$ through cross modal attention mechanism\;
    Output semantic prompts $e'_{top-K}$ by maximum similarity retrieval\;
    Connect $e'_{top-K}$ and ${\bm{\mathit{Y}}}^{t}_{ij,\text{middle}}$ and output predicted channel information $\bm{\hat{\mathit{H}}}_d^t$ through LLM backbone network and output projection\;
    Compute $\mathcal{L}_{NMSE}$ via (25)\;
    Compute $\mathcal{L}_{ce}$ via (26)\;
    Compute $\mathcal{L}_{all}$ from $\mathcal{L}_{ce}$ and $\mathcal{L}_{ctc}$ via (24)\;
    Update the weight of the model via Adam\;
}
\Output{Trained CSI-ALM}
\end{algorithm}

After completing the training of the teacher model, CSI-ALM, we proceed with training the student model. We first freeze the teacher model and obtain the attention relationship matrix from the last transformer layer of the teacher model. Simultaneously, we calculate the final predicted output and attention relations for the entire student model, and perform joint training using the following loss functions:
\begin{equation}
    \mathcal{L}_{student} =  \mathcal{L}_{NMSE} +  \lambda_2 \mathcal{L}_{attention},
\end{equation}where $\mathcal{L}_{NMSE}$ represents the CSI task prediction loss, and $\mathcal{L}_{attention}$ is the KL divergence loss of the attention relationships.

We provide a detailed description of our training method in Algorithm 1 and Algorithm 2.

\begin{algorithm}
\caption{Training Algorithm of the CSI-ALM-Light}
\label{alg:stage1}
\SetKwInOut{Input}{Input}
\SetKwInOut{Output}{Output}

\Input{Trained CSI-ALM, historical channel information ${\bm{\mathit{H}}_u^{t-1}}, {\bm{\mathit{H}}_u^{t-2}}, \dots, {\bm{\mathit{H}}_u^{t-T}}$ and true values of predicted channel information $\bm{{\mathit{H}}}_d^t$}
\While{Stop criterion is not met}{
    
    Output $\bm{R}_{xy,a}^{\text{teacher}}$ by the CSI-ALM\;
    Output $\bm{R}_{xy,a}^{\text{student}}$ and $\bm{\hat{\mathit{H}}}_d ^{t,\text{student}}$ by the CSI-ALM-Light\;
    Compute $ \mathcal{L}_{NMSE}$ via (25)\;

    Compute $\mathcal{L}_{attention}$ via (20)\;
    Compute $\mathcal{L}_{all}$ from $\mathcal{L}_{ce}$ and $\mathcal{L}_{ctc}$ via (27)\;
    Update the weight of the model via Adam\;
}
\Output{Trained CSI-ALM-Light}
\end{algorithm}

\section{NUMERICAL EXPERIMENTS}
This section presents our simulation results. First, we provide a detailed explanation of the simulation setup. Then, we compare CSI-ALM with several existing methods to evaluate the effectiveness of the proposed approach. Additionally, we conduct an ablation study to assess the benefits of leveraging pre-trained knowledge in LLMs. Finally, the proposed model's lightweight algorithm is tested, demonstrating a good balance between performance and resource overhead.
\subsection{Experimental Setup}\label{AA}
In this experiment, we utilize the channel generator QuaDRiGa\cite{b40} to generate a time-varying CSI dataset that complies with the 3GPP standard. We consider a MISO-OFDM system, where the BS is equipped with a dual-polarized uniform planar array (UPA) with \(N_h = N_v = 4\), while each user is equipped with a single omnidirectional antenna. The antenna spacing is set to half the wavelength at the center frequency. We assume that both the uplink and downlink channels have a bandwidth of 8.64 MHz, with a pilot frequency spacing of 180 kHz. 

We consider the 3GPP Urban Macro (UMa) channel model, with the initial user position randomized. The training, validation, and test sets contain 9000, 1000, and 1000 samples respectively, with user speeds uniformly distributed between 10km/h and 100km/h. In both the training and validation datasets, 10 distinct samples are generated for each velocity, while in the test dataset, each velocity is associated with 100 independent samples to ensure a more comprehensive evaluation of model generalization. For both TDD and FDD systems, the center frequency of the uplink carrier is set to 2.4GHz. In FDD systems, the carrier frequency is chosen as the band adjacent to the uplink channel.
During training, we employ the Adam optimizer with a batch size of 1024 and a learning rate of $1.0\mathrm{e}{-4}$
. During training, we employ the Adam optimizer with a batch size of 1024 and a learning rate of $1.0\mathrm{e}{-4}$.
 
 We select advanced neural network models for channel prediction, including CNN, RNN, LSTM, GRU, and Transformer, as our baseline methods. The hidden layer dimension of the RNN, LSTM, and GRU is set to 256, with a total of 3 layers. The CNN model consists of 10 convolutional layers with a kernel size of 3×3. The Transformer model uses an encoder-decoder structure with a hidden layer dimension of 384, 3 encoder layers, and 2 decoder layers. In the experiment, we directly used the CSI values of the uplink channel over the next L steps as the predicted CSI values for the downlink channel. Since our goal is to demonstrate the effectiveness of our proposed cross-modal semantic alignment method, and to balance performance and computational cost for experimental purposes, we select the smallest GPT-2 variant as the LLM backbone network in our proposed method. Its Transformer layer has a hidden dimension of 768 and the first six layers are deployed. The length of the semantic hint is set to 4. Performance improves with increasing model parameters, and our method is also applicable to other more advanced pre-trained LLMs, such as GPT-3 and LLaMA-7B. For CSI-ALM-Light, the architecture adopts a Decoder-only Transformer consisting of 3 hidden layers, each with a hidden size of 128 and 8 attention heads. The length of the soft prompt is set to 4.

\subsection{CSI Prediction Performance}\label{BB}
\begin{figure}[htbp]
    \centering
    \includegraphics[width=0.49\textwidth]{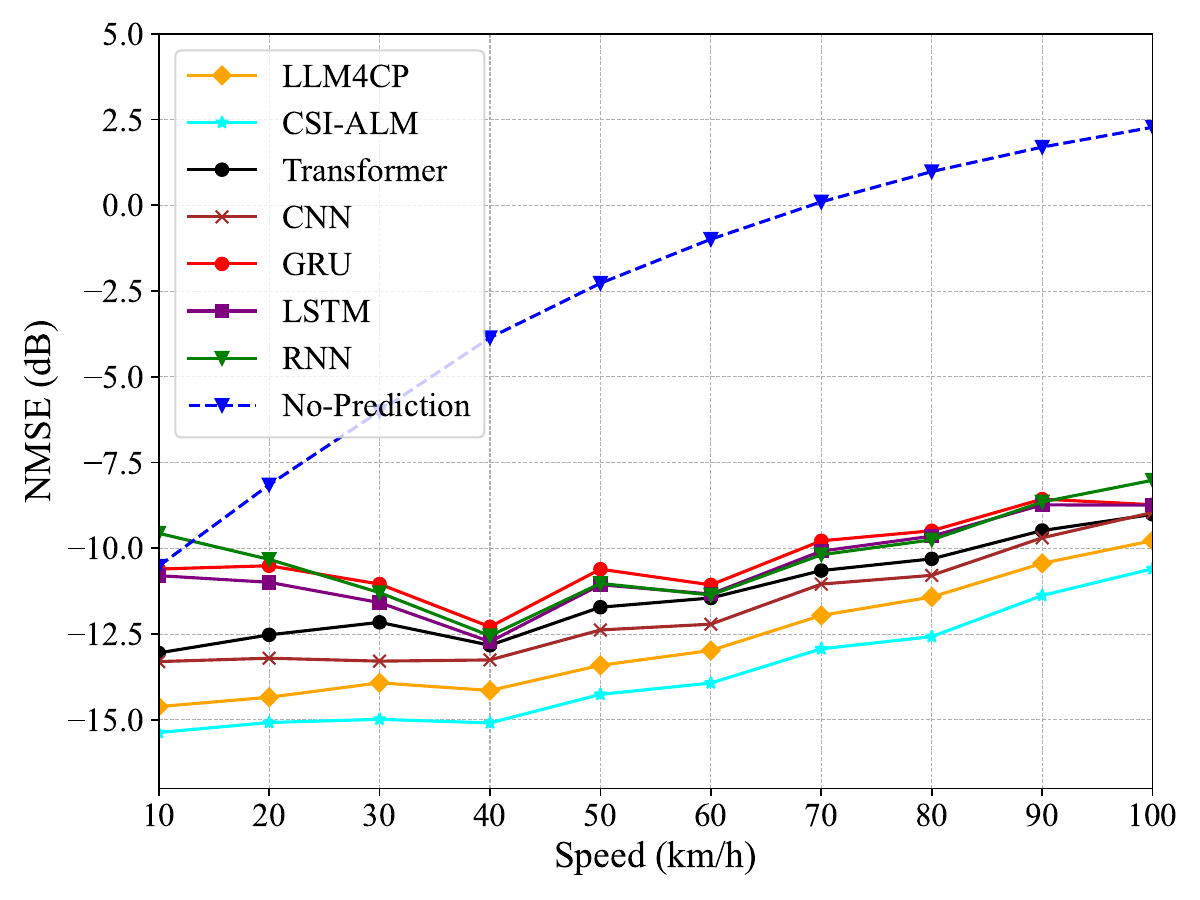} 
    \caption{The NMSE performance of CSI-ALM and other baselines versus different user velocities for TDD systems.}
    \label{fig5}
\end{figure}
\begin{figure}[htbp]
    \centering
    \includegraphics[width=0.49\textwidth]{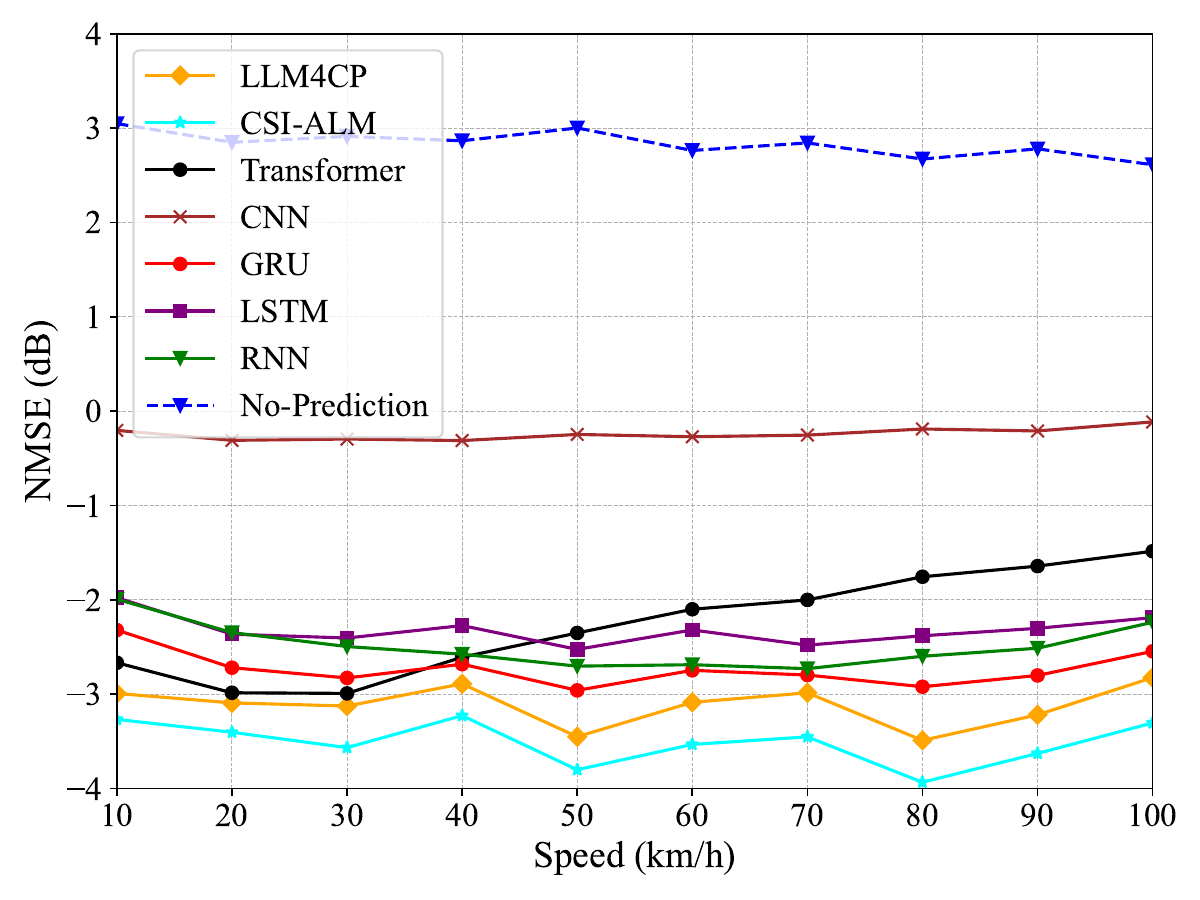} 
    \caption{The NMSE performance of CSI-ALM and other baselines versus different user velocities for FDD systems.}
    \label{fig6}
\end{figure}
In our experiments, we employ CNN\cite{b23}, RNN\cite{b24}, GRU\cite{b26}, LSTM\cite{b25}, Transformer\cite{b27}, and the existing LLM-based LLM4CP\cite{b37} as baseline methods. We first evaluate the NMSE performance of different approaches under varying user mobility speeds in TDD and FDD systems. As shown in the Fig.$\ref{fig5}$, the results in TDD system indicate that as mobility speed increases, the NMSE of all methods deteriorates. However, we observe that for recurrent models—including GRU, LSTM, and RNN—the prediction accuracy does not always degrade as user speed increases. This can be attributed to the fact that, at very low speeds, the temporal periodicity of CSI features is weak, making it difficult for these models to capture meaningful temporal dependencies. Conversely, at higher speeds, the coherence time of the wireless channel decreases, leading to more rapid channel variations that also challenge the temporal modeling capability of these methods. However, due to the superior feature extraction capability of LLMs, LLM-based methods significantly outperform traditional deep learning approaches. Moreover, our approach introduces a modality alignment mechanism specifically designed to bridge the substantial gap between CSI features and linguistic knowledge, thereby further enhancing the potential of LLMs in CSI prediction tasks. Compared to LLM4CP, our approach achieves even higher prediction accuracy with further reduced NMSE, as modality alignment ensures that CSI representations are aligned with the linguistic feature space, allowing pretrained linguistic knowledge to be effectively leveraged for communication tasks. For FDD systems, we can see that the pattern of experimental results is roughly similar to that of TDD systems. However, due to the more complex channel characteristics caused by the non reciprocal uplink and downlink channels in FDD systems, the prediction performance of each method has decreased. Nevertheless, our proposed method still achieved the best results. Additionally, our analysis reveals that approaches leveraging LLM exhibit enhanced advantages, which can be attributed to their superior sequence modeling capabilities in capturing temporal dependencies and noise-robust feature extraction within dynamic channel environments.

\begin{figure}[htbp]
    \centering
    \includegraphics[width=0.49\textwidth]{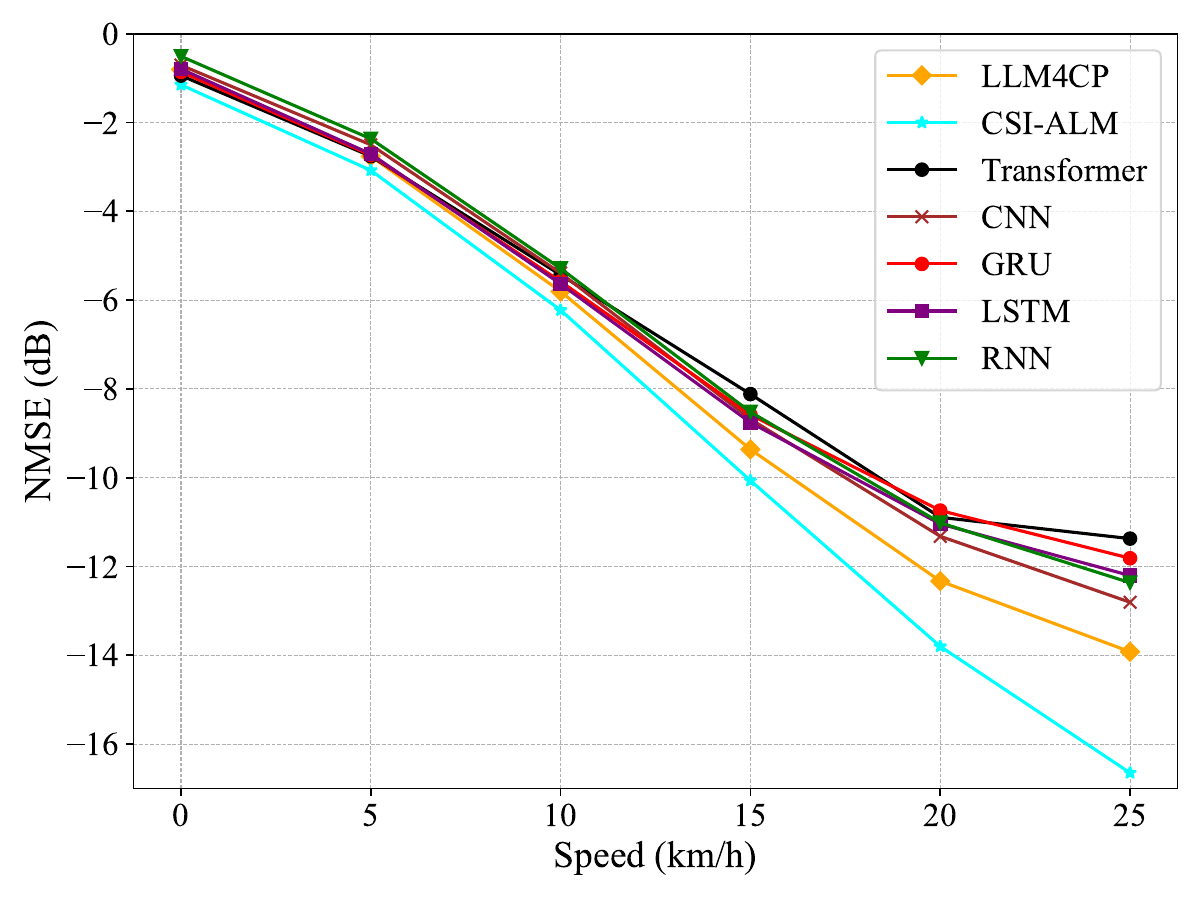} 
    \caption{The NMSE performance versus SNR of nosiy historical CSI for TDD systems.}
    \label{fig7}
\end{figure}
\begin{figure}[htbp]
    \centering
    \includegraphics[width=0.49\textwidth]{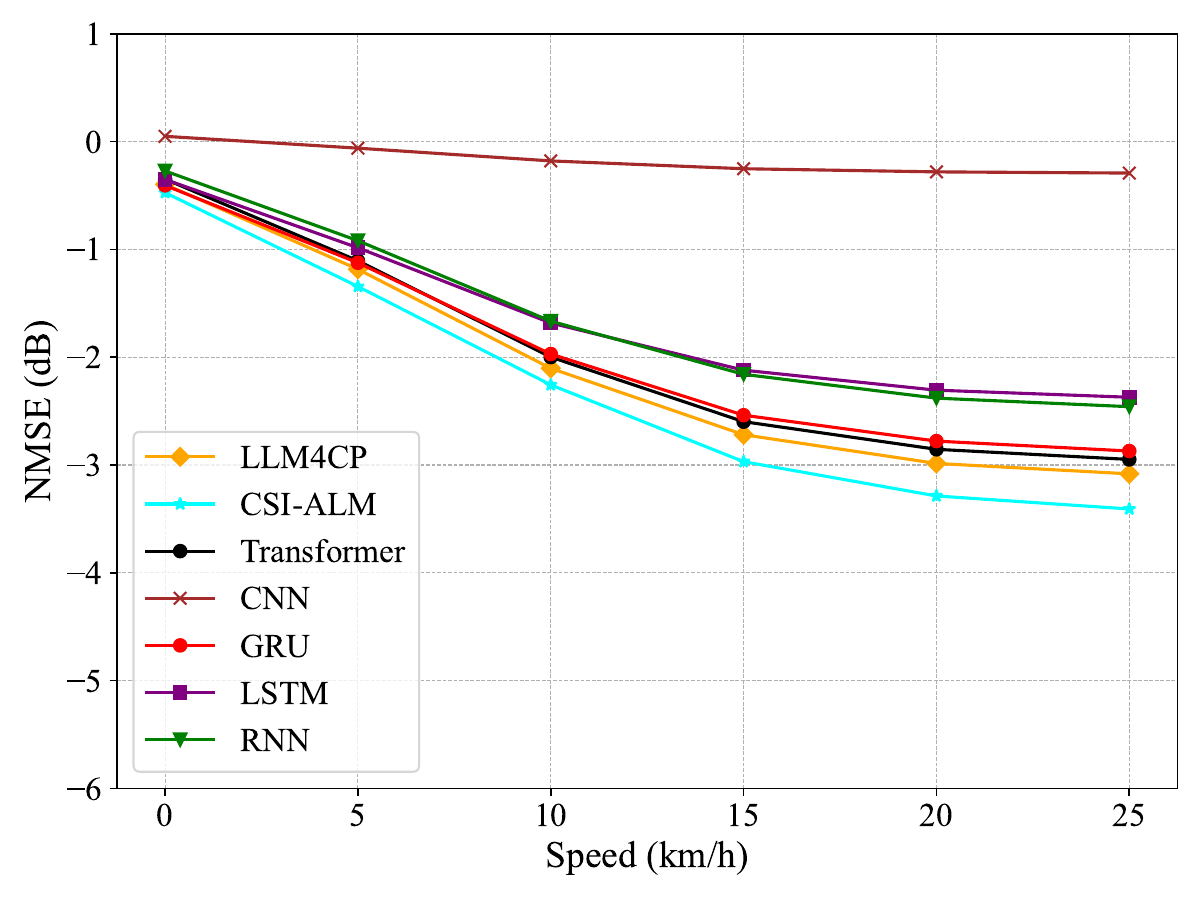} 
    \caption{The NMSE performance versus SNR of nosiy historical CSI for FDD systems.}
    \label{fig8}
\end{figure}

\begin{table*}[ht]
\centering
\caption{Average NMSE performance of each method in TDD and FDD scenarios.}
\label{TableI}
\begin{tabular}{|c|c|c|c|c|c|c|c|}
\hline

\textbf{Scenario} & \textbf{CSI-ALM} & \textbf{LLM4CP} & \textbf{Transformer}& \textbf{CNN} & \textbf{GRU}& \textbf{LSTM}& \textbf{RNN}\\
\hline
TDD & -13.62dB & -12.61dB & -11.32dB & -11.81dB & -10.27dB & -10.59dB & -10.27dB\\
\hline
FDD & -3.51dB & -3.09dB & -2.25dB & -0.24dB & -2.73dB & -2.32dB & -2.48dB\\
\hline

\end{tabular}
\end{table*}

Due to channel estimation errors in practical environments, testing the robustness of models under noisy conditions is crucial. Therefore, in the scenario of a moving speed of 30km/h, we add noise with different SNR to the historical channel information input of the model to evaluate its robustness to noise. We present the CSI prediction results for TDD and FDD systems at different SNRs in Fig.$\ref{fig7}$ and Fig.$\ref{fig8}$. As shown in the figures, all methods exhibit poor performance at low SNRs in both TDD and FDD scenarios, and the prediction errors of all methods gradually decrease as the SNR increases. However, the proposed CSI-LLM method not only achieves the best performance at all SNRs but also shows the smoothest variation with changing SNRs, indicating that our method has high robustness. This is critical for robust channel prediction in real-world scenarios.

\begin{figure}[htbp]
    \centering
    \includegraphics[width=0.49\textwidth]{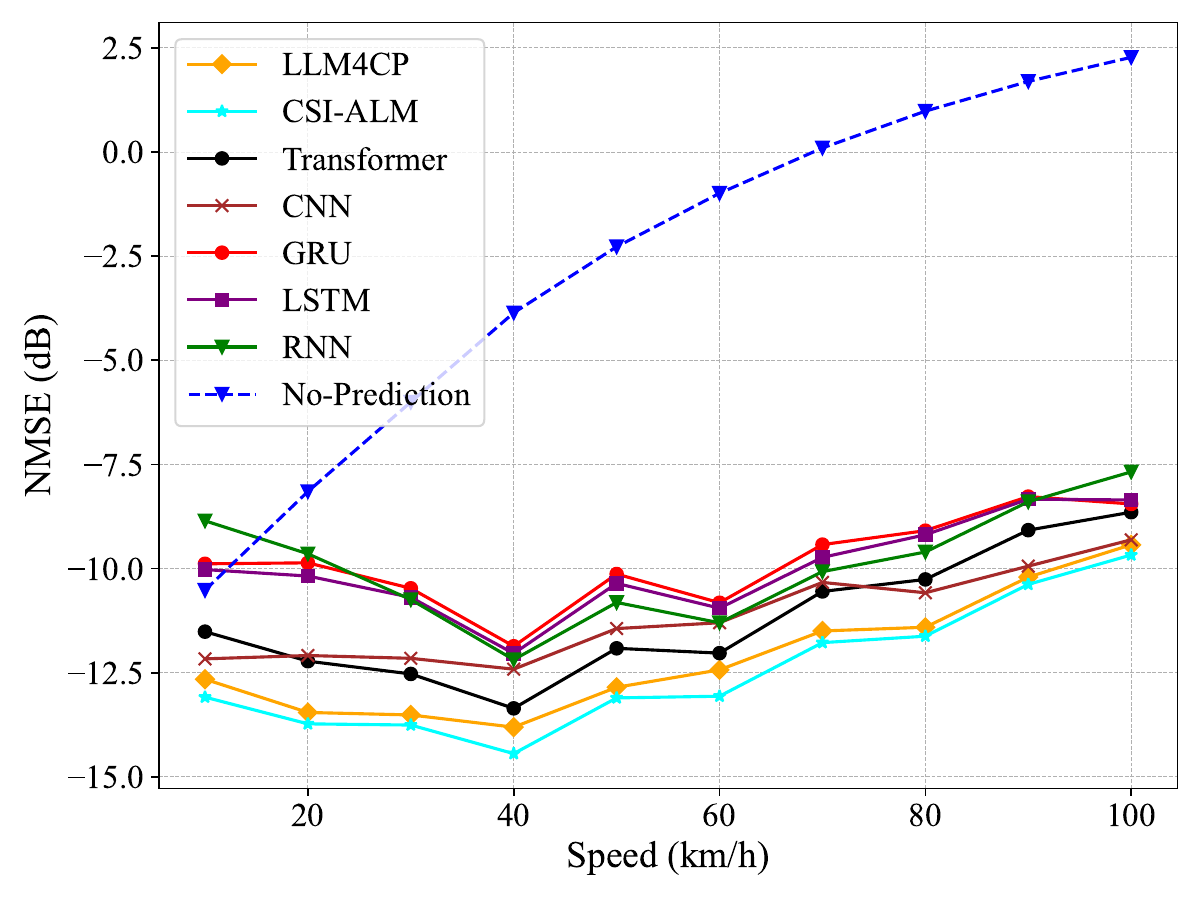} 
    \caption{The NMSE performance of CSI-ALM and other baselines versus different user velocities for TDD systems (Few-shot: 10\% training dataset).}
    \label{fig9}
\end{figure}

\begin{figure}[htbp]
    \centering
    \includegraphics[width=0.49\textwidth]{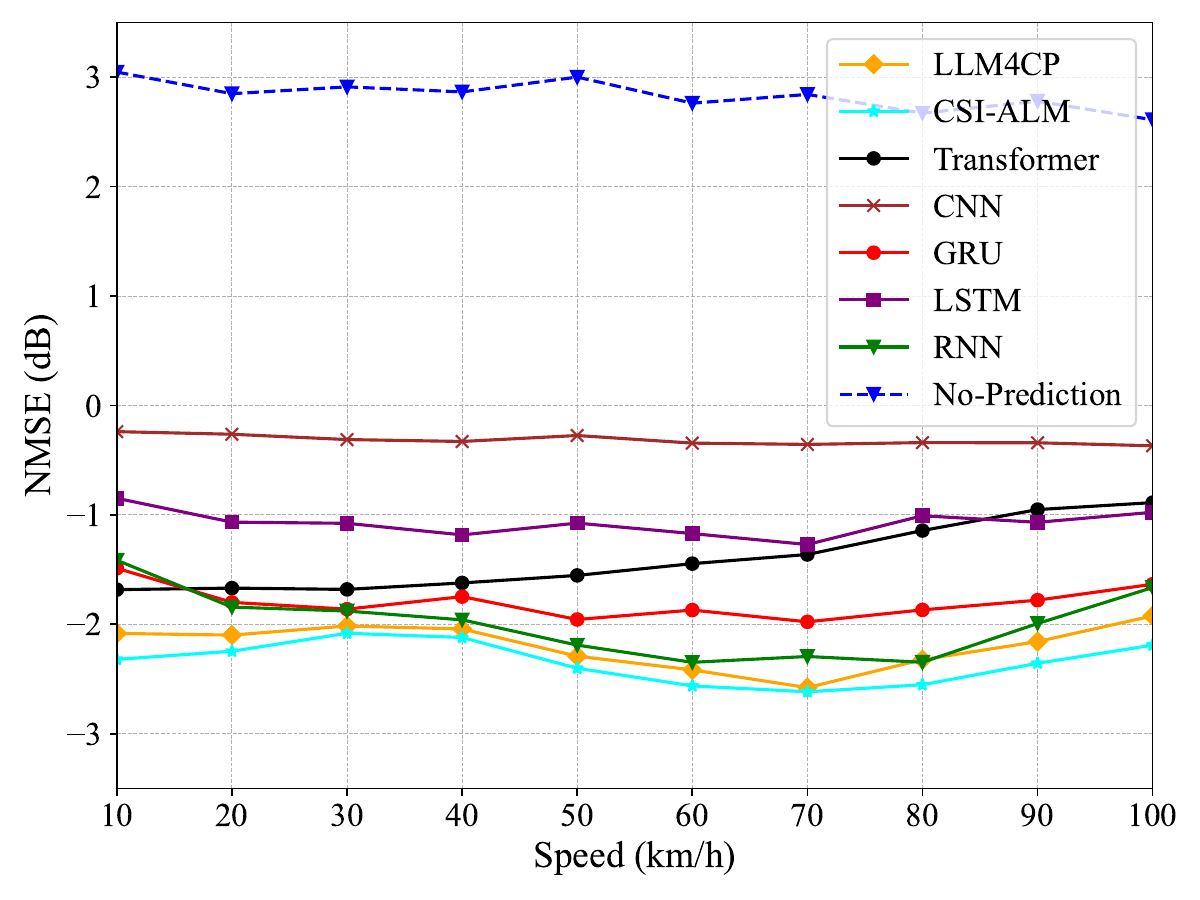} 
    \caption{The NMSE performance of CSI-ALM and other baselines versus different user velocities for FDD systems (Few-shot: 10\% training dataset).}
    \label{fig10}
\end{figure}

In practical applications, few-shot prediction is crucial for the rapid deployment of deep learning-based models. To evaluate the few-shot learning capability of our CSI-LLM, we conduct experiments in TDD and FDD scenarios, where only 10\% of the training data is used for training, and then assess the models' performance on the test set. As show in Fig.$\ref{fig9}$, in TDD scenario, the results show that under limited data conditions, the feature extraction capability of all methods degrades. However, LLM-based approaches, benefiting from rich pretrained knowledge, achieve significantly better results. Furthermore, due to our modality alignment design, CSI features are more effectively integrated into the LLM backbone, enhancing the utilization of pretrained linguistic knowledge in CSI prediction. In the FDD scenario, as shown in Fig.$\ref{fig10}$, we can observe that due to the more complex frequency characteristics, the channel variation in the FDD scenario exhibits greater fluctuations. All methods show some performance degradation. However, thanks to the powerful sequence modeling capability of LLM and the efficient cross-modal alignment mechanism we designed, the proposed method still achieves the best performance.

\begin{figure}[htbp]
    \centering
    \includegraphics[width=0.49\textwidth]{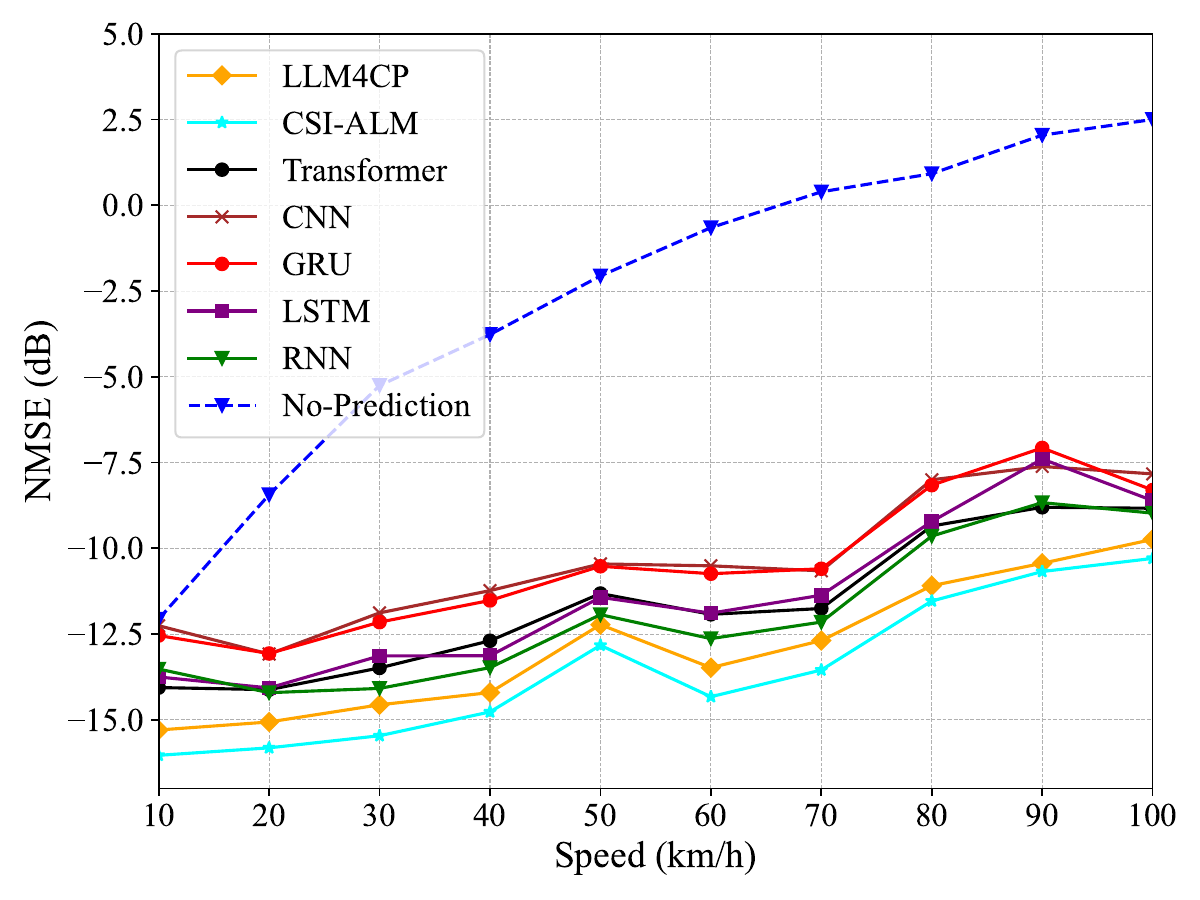} 
    \caption{The zero-shot generalization performance for the TDD systems in UMi scenario.}
    \label{fig11}
\end{figure}
\begin{figure}[htbp]
    \centering
    \includegraphics[width=0.49\textwidth]{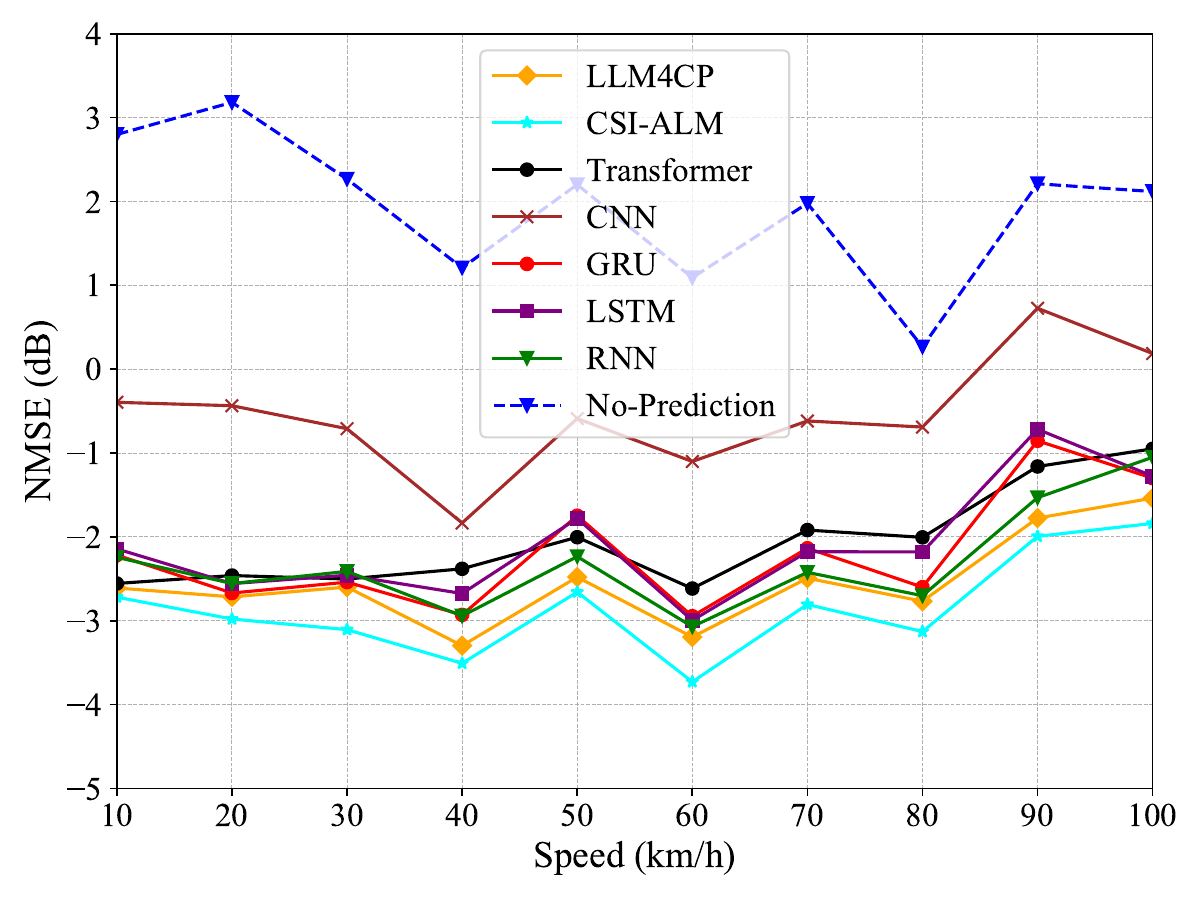} 
    \caption{The zero-shot generalization performance for the FDD systems in UMi scenario.}
    \label{fig12}
\end{figure}

\begin{figure}[htbp]
    \centering
    \includegraphics[width=0.49\textwidth]{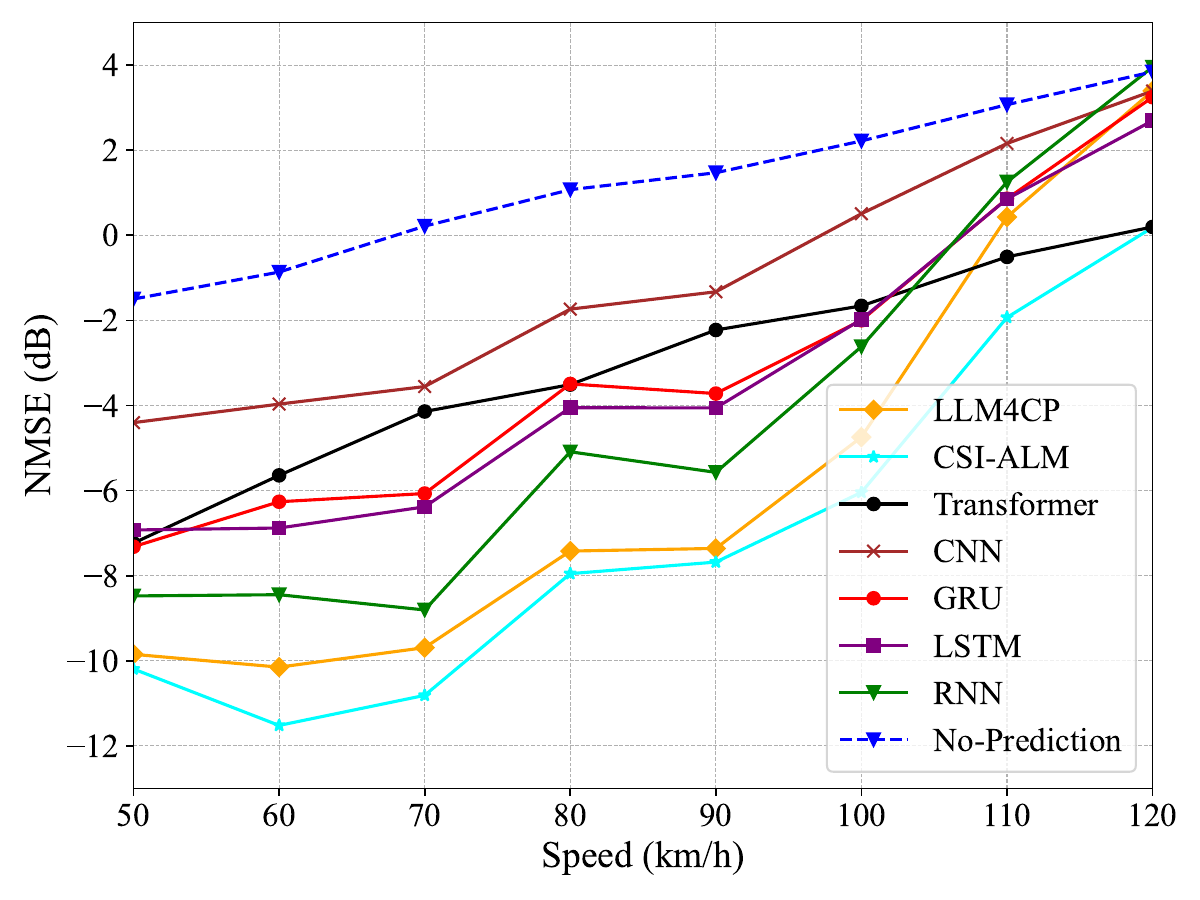} 
    \caption{The zero-shot generalization performance for the FDD systems in highway scenario.}
    \label{fig13}
\end{figure}
Additionally, considering TDD system, we evaluate the zero-shot Compared to the 3GPP Uma channel scenario, the Umi scenario has more complex environmental characteristics. The Umi scenario has a smaller coverage area and higher user density, with base stations deployed at lower heights. Additionally, the channel fading and multipath effects in the Umi scenario are more pronounced, and issues such as co-channel interference and neighboring cell interference are more prominent. Therefore, we evaluate the zero-shot generalization capability of all models by training them in the Uma scenario and directly applying them to the Umi scenario, while keeping all other settings unchanged. The results in Fig.$\ref{fig11}$ show that, during model transfer testing in the TDD system, the LLM-based method retains the strong generalization capability observed in NLP tasks. Furthermore, our CSI-ALM framework, equipped with modality alignment, further improves cross-scenario prediction performance, outperforming other methods. We then conducted the same transfer testing in the FDD system. As shown in Fig.$\ref{fig12}$, due to the strong frequency selectivity of the channel, all methods exhibit poor NMSE performance. However, our proposed approach still achieves the smallest NMSE, significantly outperforming other methods. This is because our proposed modality alignment scheme can better leverage the pre-trained knowledge of LLM and align the features between textual knowledge and CSI data with a small amount of training data. To more clearly demonstrate the numerical results, we also present the average NSME values of each model under both TDD and FDD scenarios in Table $\ref{TableI}$. As shown in Table $\ref{TableI}$, in TDD scenarios the proposed CSI-ALM method achieves an average gain of over 1 dB compared with the state-of-the-art LLM4CP method. In the FDD scenario, despite the increased channel complexity and the greater difficulty of CSI feature extraction, the proposed method still achieves an average performance gain of about 0.4 dB, demonstrating its robustness under challenging conditions.
 
Finally, we further evaluate the generalization capability of the proposed method under high-mobility communication scenarios. In such FDD environments, the pronounced Doppler effect often causes conventional algorithms to suffer from a rapid performance degradation. As illustrated in the Fig.~\ref{fig13}, our proposed method demonstrates stronger robustness in this setting, offering an effective solution for channel prediction in extreme communication environments.

\begin{table*}[ht]
\centering
\caption{Inference speed and complexity comparison.}
\label{TableII}
\begin{tabular}{|c|c|c|c|}
\hline

\textbf{method} & \textbf{Training parameter} & \textbf{Model parameter} & \textbf{inference time} \\
\hline
GRU & 1.49M & 1.49M & 7.74ms \\
\hline
LSTM & 1.96M & 1.96M & 8.34ms \\
\hline
RNN & 0.51M & 0.51M & 6.72ms \\
\hline
Transformer & 6.78M & 6.78M & 80.9ms \\
\hline
CNN & 3.14M & 3.14M & 1.38ms \\
\hline
LLM4CP & 1.77M & 83.6M & 20.1ms \\
\hline
CSI-ALM-Light & 0.349M & 0.349M & 2.41ms \\
\hline
CSI-ALM & 11.8M & 93.8M & 198.2ms \\
\hline
\end{tabular}
\end{table*} 
\subsection{Ablation Experiment}\label{CC}

\begin{figure}[htbp]
    \centering
    \includegraphics[width=0.49\textwidth]{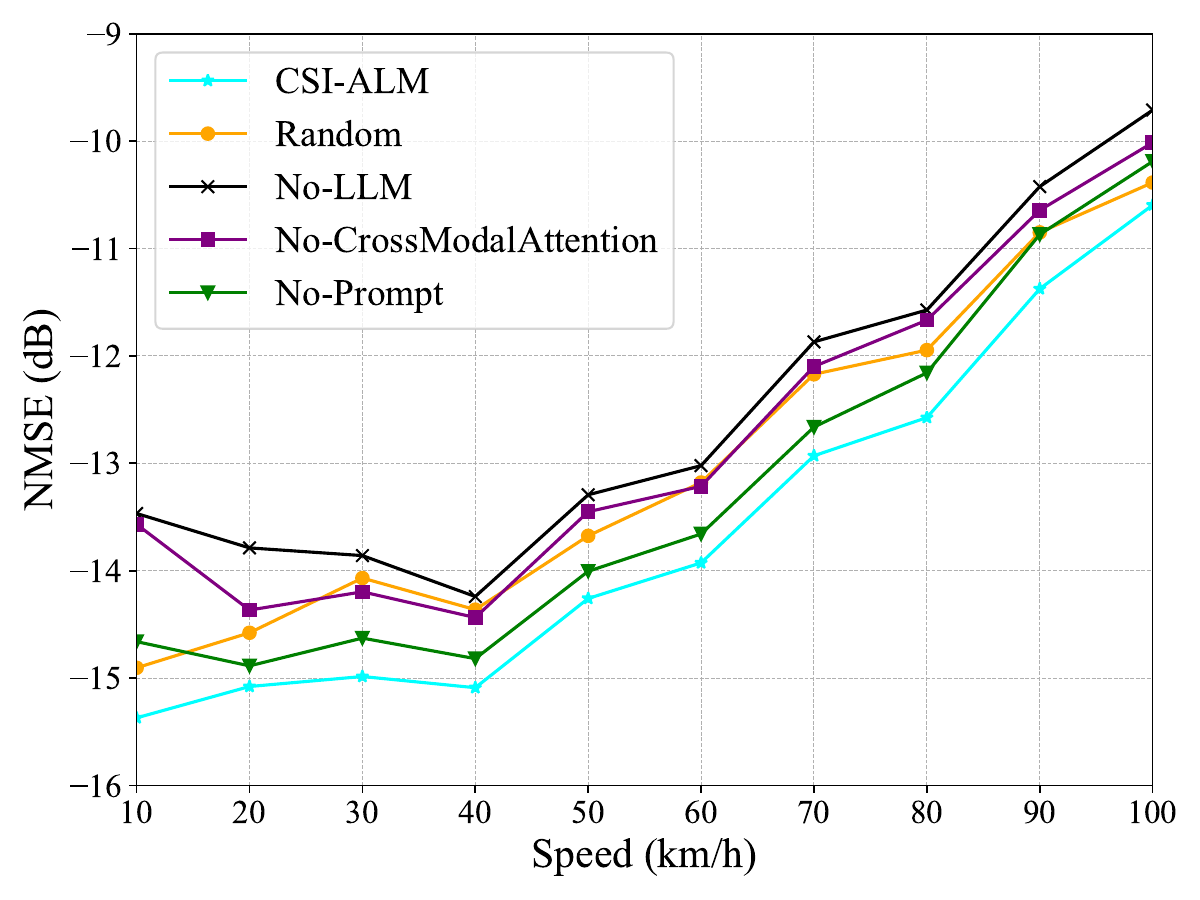} 
    \caption{Results of ablation experiment.}
    \label{fig14}
\end{figure}

In this section, we conduct extensive ablation experiments to verify the effectiveness of each module in our CSI-ALM. The fundamental advantage of LLM-based CSI prediction over traditional methods lies in its ability to leverage pretrained knowledge. To further analyze this aspect, we compare three different initialization strategies: (1) random initialization, (2) initialization from a pretrained base model, and (3) removal of the LLM backbone. We conducted ablation experiments under TDD system, as illustrated in Fig.~\ref{fig14}, initializing with a pretrained base model yields the best performance, significantly outperforming the other two approaches. The randomly initialized model lacks access to pretrained knowledge, rendering the modality alignment mechanism ineffective. Consequently, its performance becomes comparable to that of the model without the LLM backbone. These results further confirm the critical role of pretrained knowledge in enhancing the effectiveness of channel prediction.

Moreover, as shown in Fig.~\ref{fig14}, emoving the cross-modal attention module and semantic prompts, both designed to align CSI features with textual knowledge, leads to a significant degradation in channel prediction performance. Among these two factors, the impact of removing the cross-modal attention module is particularly pronounced. In fact, without this module, performance drops to a level comparable to that of randomly initialized LLM backbones. This observation further demonstrates that, without effective modality alignment, the pretrained knowledge embedded in LLMs cannot be fully leveraged, making it difficult to enhance channel prediction performance.
\begin{table}[ht]
\centering
\caption{NMSE performance of CSI-ALM-light with different model sizes.}
\label{TableIII}
\begin{tabular}{|c|c|c|c|}
\hline

 \textbf{NMSE} & \textbf{Parameter} & \textbf{inference time} \\
\hline
  -11.19dB & 0.349M & 2.41ms \\
\hline
  -11.42dB & 0.449M & 2.46ms \\
\hline
  -11.64dB & 0.548M & 2.52ms \\
\hline
  -11.91dB & 0.648M & 2.84ms \\
\hline

\end{tabular}
\end{table}
\subsection{Model Lightweighting Experiment}
\begin{figure}[htbp]
    \centering
    \includegraphics[width=0.49\textwidth]{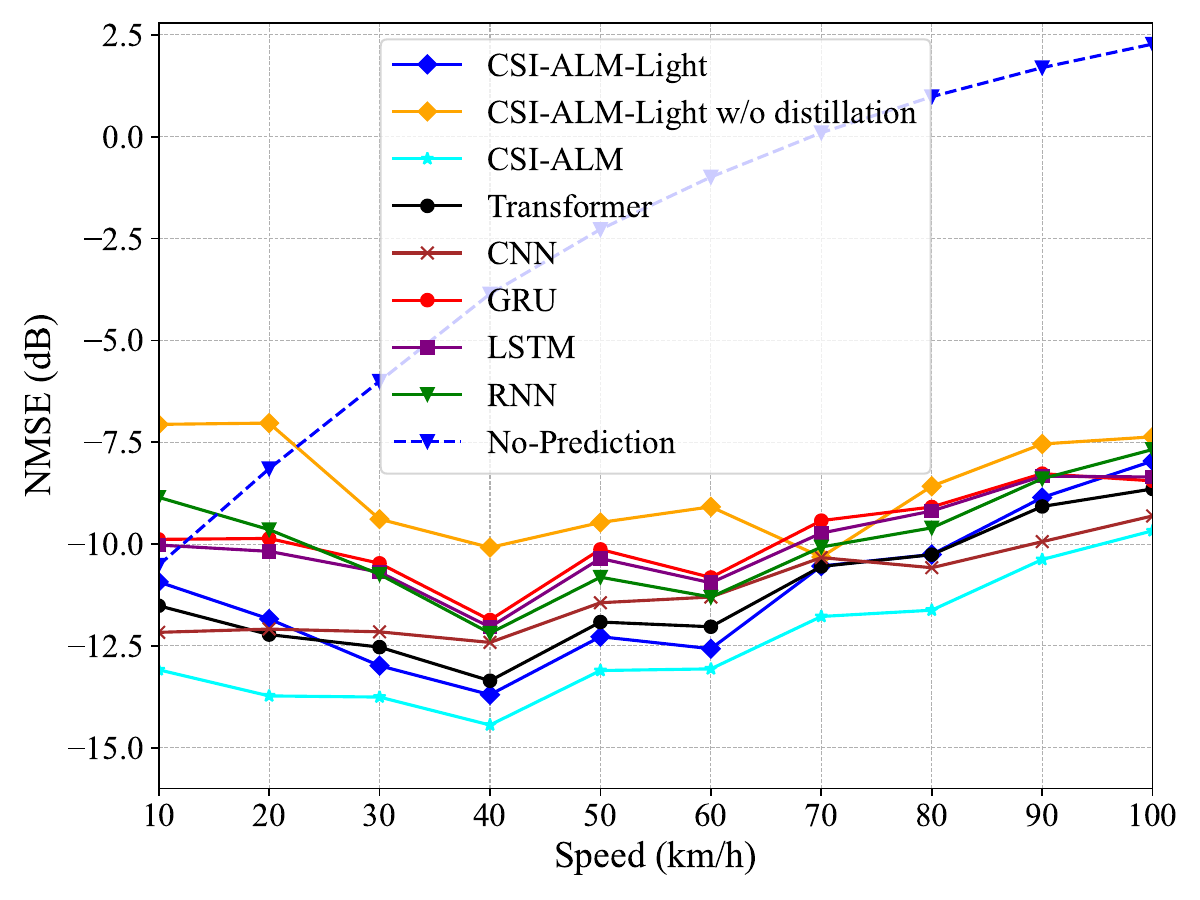} 
    \caption{Results of lightweighting experiment in TDD systems.}
    \label{fig15}
\end{figure}

\begin{figure}[htbp]
    \centering
    \includegraphics[width=0.49\textwidth]{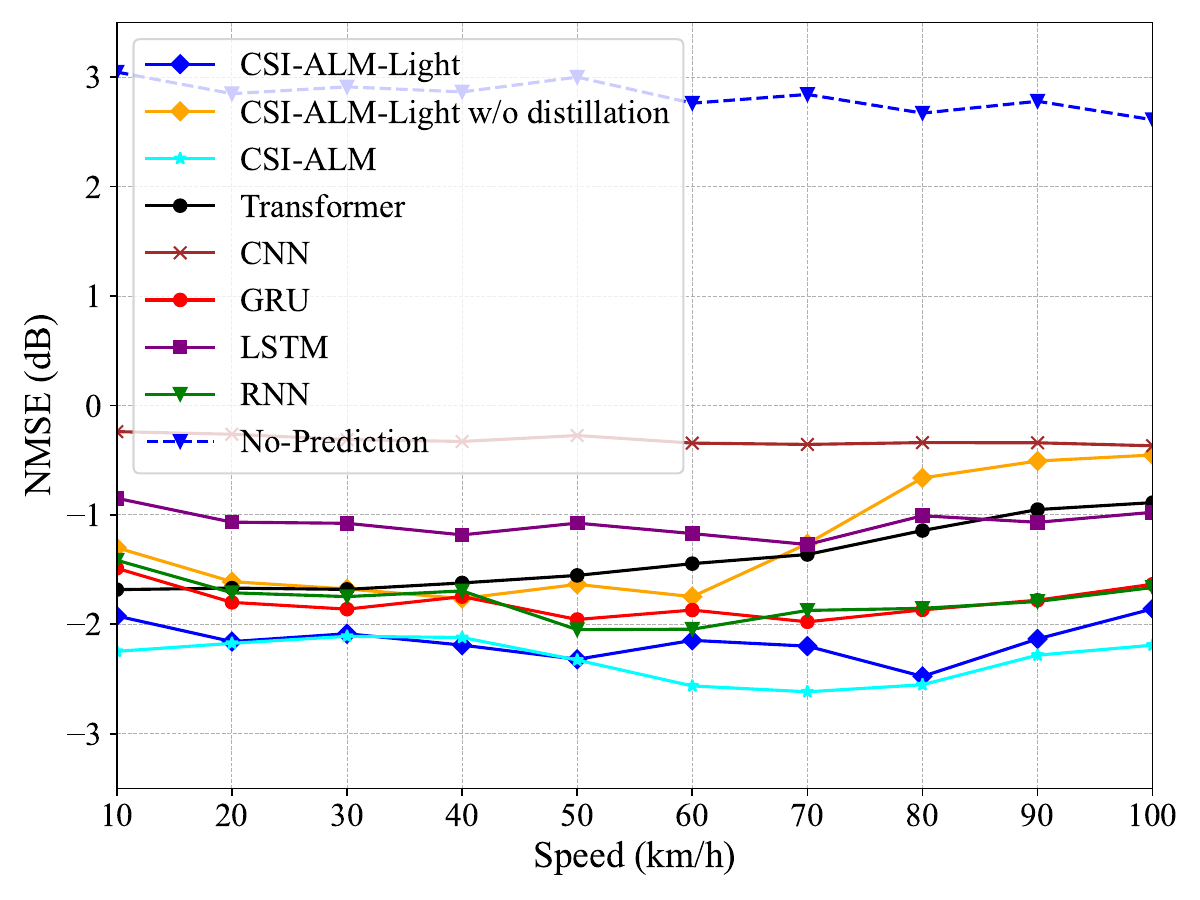} 
    \caption{Results of lightweighting experiment in FDD systems.}
    \label{fig16}
\end{figure}

In this section, we conduct experiments on the deployment of lightweight models. Specifically, we use a knowledge distillation method based on self attention relationships to obtain CSI-ALM-Light, whose parameters are similar to those of a small deep learning model. We train the model using 10\% of the original training set and evaluate the predictive performance of CSI-ALM-Light and other small deep learning models under TDD and FDD systems. As shown in Fig. $\ref{fig15} $ and $\ref {fig16} $, When the amount of training data is small, in most speed scenarios, although our CSI-ALM-Light has a simple structure, its NMSE performance in predicting CSI is still better than other traditional deep learning methods. In addition, we observe a very interesting phenomenon: when only 10\% of the training data is used, the performance of our student model was almost the same as that of the teacher model. This is because, in the case of limited data, CSI-ALM often cannot learn the general patterns of the data, while CSI-ALM-Light, guided by well-trained CSI-ALM, can learn richer prior knowledge and store it in the attention layer. By refining the self attention relationship, the CSI-ALM-Light can quickly learn the prior knowledge of the CSI-ALM, thereby achieving better performance. At the same time, we present the performance of training the student model separately without knowledge distillation. It can be seen that the predictive performance of training the student model separately is far inferior to that of the CSI-ALM-Light that has undergone knowledge distillation, whether in TDD or FDD scenarios. This further demonstrates the effectiveness of our proposed knowledge distillation method based on self attention relationships.

At the same time, although CSI-ALM-Light performs worse than baseline methods under certain speed conditions in TDD scenarios, this is mainly due to model capacity limitations. We also evaluate CSI-ALM-Light of model different sizes in TDD scenarios. As shown in the Table $\ref{TableII}$, even a small increase in model capacity leads to a significant improvement in performance. This implies the model can be flexibly scaled according to system requirements to balance deployment efficiency and prediction accuracy.
\subsection{Modal Alignment Validity Analysis}

\begin{figure}[htbp]
    \centering
    \includegraphics[width=0.4\textwidth]{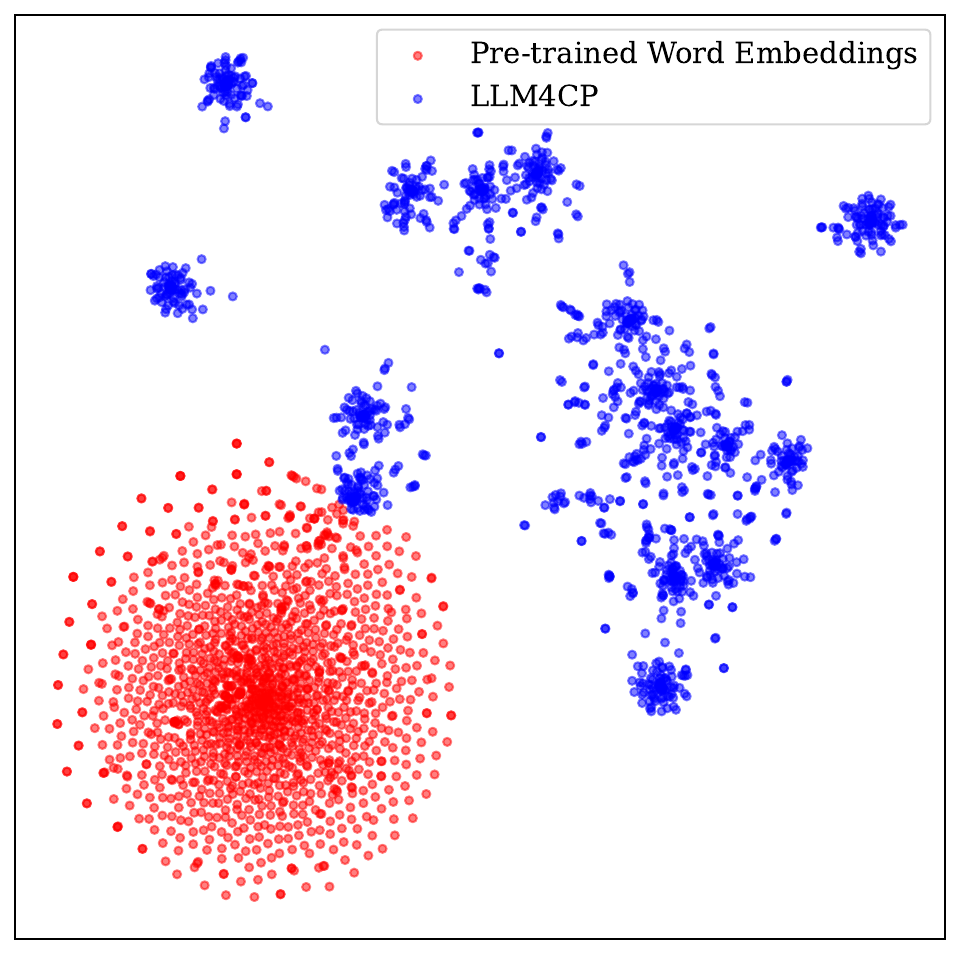} 
    \caption{t-SNE visualization of LLM4CP pre-trained word Embeddings and CSI Embeddings.}
    \label{fig17}
\end{figure}

\begin{figure}[htbp]
    \centering
    \includegraphics[width=0.4\textwidth]{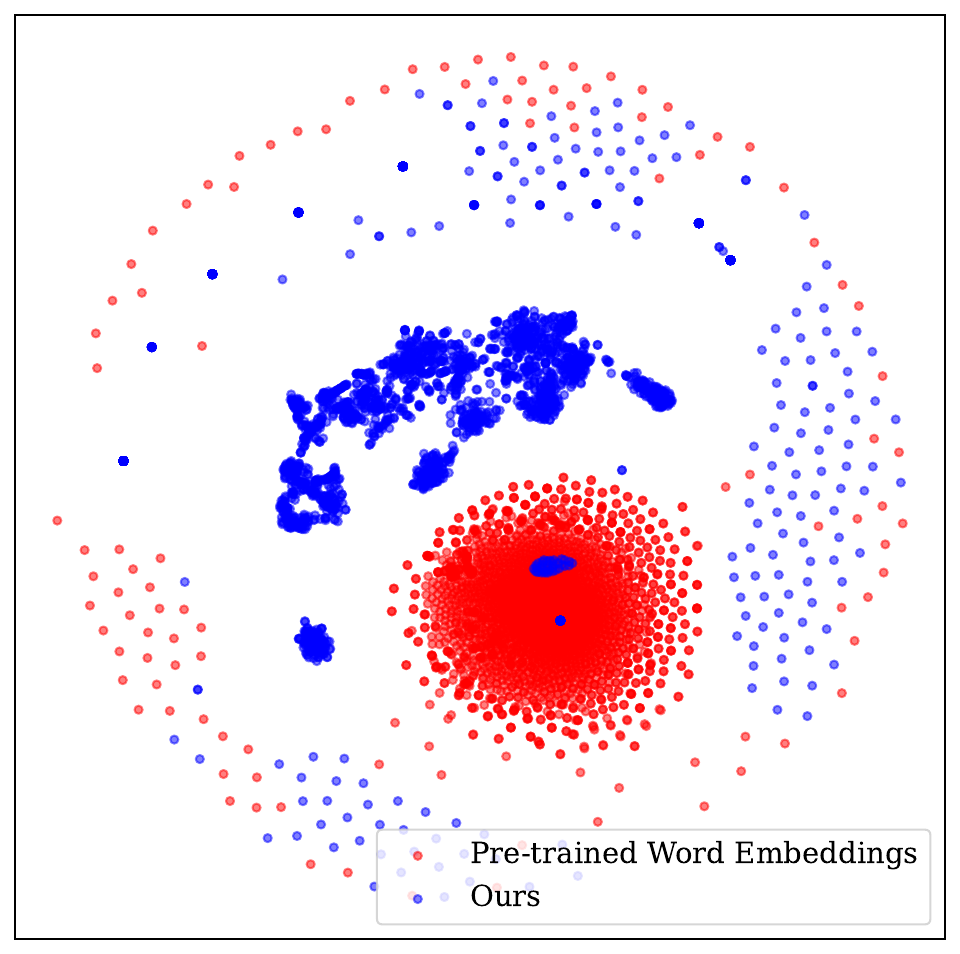} 
    \caption{t-SNE visualization of CSI-ALM pre-trained word Embeddings and CSI Embeddings.}
    \label{fig18}
\end{figure}

In this section, we provide a thorough analysis of the effectiveness of our proposed modality alignment method. As mentioned earlier, the pre-trained LLM inherently operates within the linguistic semantic space, as it is trained entirely on a text corpus. Therefore, when directly applied to physical domain representations such as CSI, it fails to effectively leverage its rich semantic priors, resulting in a significant modality gap. To address this issue, we introduce a modality alignment strategy based on a cross-modal attention mechanism, enabling the model to project CSI features into a shared latent semantic space compatible with the LLM pre-trained linguistic representations. To further validate the effectiveness of this mechanism, we visualize the feature distributions of CSI embeddings and pre-trained word embeddings for LLM4CP and our proposed CSI-ALM, as shown in Fig. $\ref{fig17} $ and $\ref{fig18} $. The results clearly reveal fundamental differences in representational behavior. Specifically, the CSI embeddings generated by LLM4CP tend to form multiple isolated clusters that are significantly separated from the word embeddings. This scattered distribution indicates that LLM4CP fails to establish effective alignment between CSI embeddings and the semantic space of the pre-trained language model, resulting in suboptimal knowledge transfer and, in turn, degraded prediction performance. In contrast, after applying the proposed modality alignment strategy, the CSI embeddings generated by CSI-ALM have a distribution closer to the pre-trained word embeddings and exhibit stronger structural cohesion. The emergence of these compact and semantically related clusters indicates that the CSI representations are effectively aligned with the pre-trained language feature space. This improved cohesion enables the LLM backbone to interpret CSI features in a manner consistent with its learned language structure, thereby enhancing semantic understanding and facilitating cross-domain knowledge transfer. Furthermore, the increased density of aligned embeddings reflects a reduction in modality discrepancy, implying that the model learns a more unified and semantically meaningful latent representation space. By bridging the gap between physical-domain CSI data and language semantics, our approach enables the pre-trained LLM to leverage its prior knowledge for channel prediction tasks, thereby improving the model's interpretability, robustness, and generalization across diverse communication environments.

\subsection{Analysis of the Effectiveness of Knowledge Distillation}
\begin{figure}[htbp]
    \centering
    \includegraphics[width=0.4\textwidth]{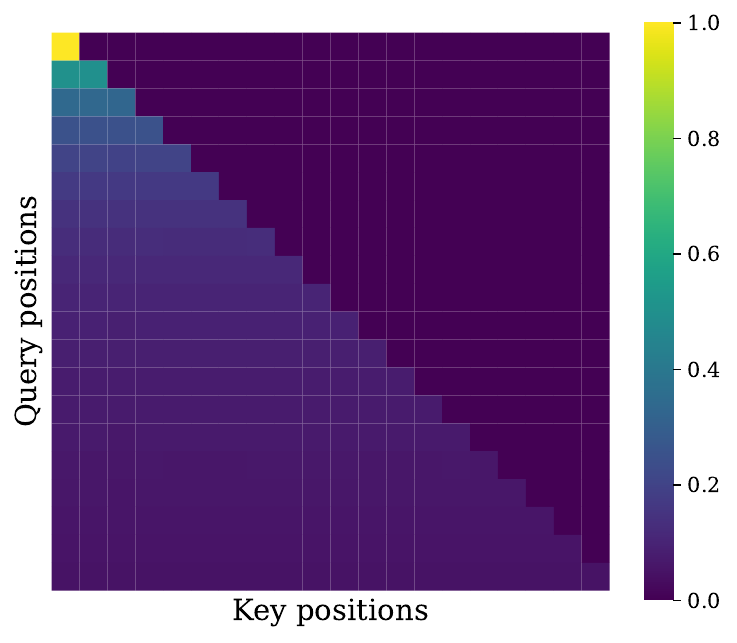} 
    \caption{Attention heatmap of the CSI-ALM.}
    \label{fig19}
\end{figure}

\begin{figure}[htbp]
    \centering
    \includegraphics[width=0.4\textwidth]{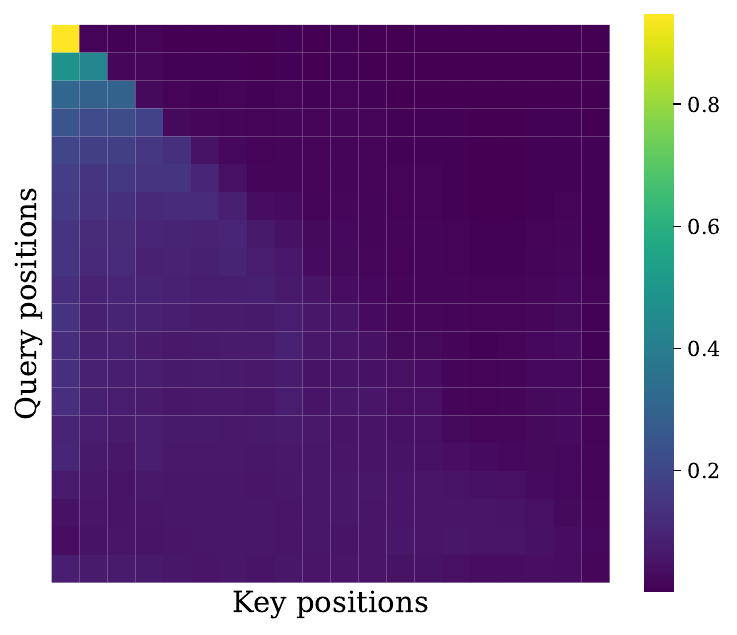} 
    \caption{Attention heatmap of the CSI-ALM-light.}
    \label{fig20}
\end{figure}

\begin{figure}[htbp]
    \centering
    \includegraphics[width=0.4\textwidth]{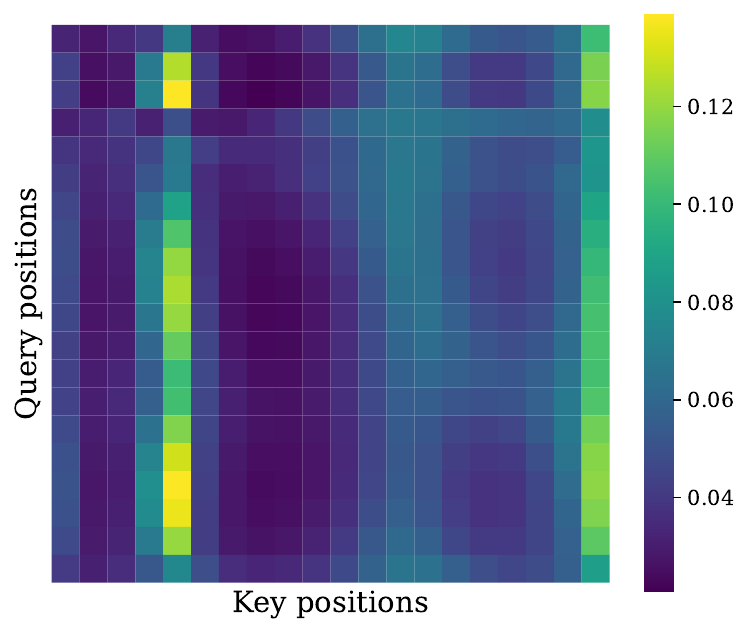} 
    \caption{Attention heatmap of the CSI-ALM-light without distillation.}
    \label{fig21}
\end{figure}

In this section, we provide a more detailed analysis to further demonstrate the effectiveness of the proposed knowledge distillation strategy. Specifically, we visualize the attention heatmaps from the last layer of CSI-ALM, CSI-ALM-light, and CSI-ALM-light without knowledge distillation. As shown in Fig. $\ref{fig19} $ and $\ref{fig20} $, the attention patterns of CSI-ALM and CSI-ALM-light exhibit a high degree of similarity, indicating that the CSI-ALM-light successfully inherits the structural knowledge captured by the CSI-ALM. This resemblance suggests that the distillation process effectively preserves the key contextual dependencies learned by the teacher model, enabling the lightweight model to maintain comparable interpretability while significantly reducing computational cost. In constrast, as shown in Fig. $\ref{fig21} $, the attention patterns of the model trained without knowledge distillation appear disordered and unstructured. These results demonstrate that the proposed distillation framework not only compresses the model but also faithfully preserves critical attention mechanisms, thereby validating the practical applicability of CSI-ALM-light in real-world channel prediction tasks.

\subsection{Training and Inference Cost}
We also compare the parameter counts and inference times of different models to evaluate their deployment challenges in practical scenarios. Our experiment was conducted on a single RTXA6000. As shown in Table $\ref{TableII}$, it can be observed that although LLM-based approaches inherently exhibit a larger parameter count due to their architectural characteristics, their inference speed remains competitive with smaller deep learning models owing to specialized acceleration mechanisms. Furthermore, the distilled student model addresses the challenge of excessive parameter size and achieves comprehensive optimization in both parameter efficiency and inference latency.

\section{Conclusion}
In this paper, we proposed a LLM channel prediction method based on semantic prompting and cross-modal alignment, named CSI-ALM. CSI-ALM first obtains the feature representation of CSI through a channel embedding module, and then combines pre-trained word embeddings with CSI features through a cross-modal attention mechanism to obtain the aligned CSI features. We also introduced a semantic prompting method that retrieves CSI features and pre-trained linguistic knowledge based on maximum similarity, further enhancing the utilization of the LLM's pre-trained knowledge. Extensive experimental results demonstrate that CSI-ALM consistently outperforms state-of-the-art deep learning methods across various communication scenarios, achieving substantial performance gains. Furthermore, our experiments showed that the proposed model lightweighting approach enabled CSI-ALM-Light, with only 0.34M parameters, to achieve performance comparable to CSI-ALM under limited training data conditions—where all models were trained using only 10\% of the original training dataset—and significantly outperformed conventional deep learning methods,  which is crucial for real-world deployment with limited training data or when applied to new environments.

\end{document}